\documentclass[12pt]{article}

\usepackage{amssymb}
\usepackage{amsmath}
\usepackage[dvipdfmx]{graphicx}
\usepackage{amsthm}   
\usepackage{authblk}   
\usepackage{pdflscape}   
\usepackage{url}
\usepackage{multirow}
\usepackage{here}
\usepackage[super]{cite}
\usepackage{bm}
\usepackage{color}

\usepackage{siunitx}
\usepackage{comment}

\usepackage{setspace}
\makeatletter

\@addtoreset{equation}{section}
\makeatother

\makeatletter
\def\mojiparline#1{
    \newcounter{mpl}
    \setcounter{mpl}{#1}
    \@tempdima=\linewidth
    \advance\@tempdima by-\value{mpl}zw
    \addtocounter{mpl}{-1}
    \divide\@tempdima by \value{mpl}
    \advance\kanjiskip by\@tempdima
    \advance\parindent by\@tempdima
}
\makeatother
\def\linesparpage#1{
    \baselineskip=\textheight
    \divide\baselineskip by #1
}

\newtheorem{remark}{Remark}

\allowdisplaybreaks[4]

\newcommand{\indep}{\mathop{\perp\!\!\!\perp}}

\newcommand{\bld}{\boldsymbol}

\usepackage[top=1in,bottom=1in,left=1in,right=1in]{geometry}

\title{Generalized Bayesian Inference for Causal Effects using Covariate Balancing Procedure}

\author[1]{Shunichiro Orihara \thanks{Address:\ 6-1-1 Shinjuku, Shinjuku-ku, Tokyo 160-8402, Japan;\ \  Email:\ orihara@tokyo-med.ac.jp}}
\author[2]{Tomotaka Momozaki}
\author[3,4]{Tomoyuki Nakagawa}

\affil[1]{Department of Health Data Science, Tokyo Medical University, Tokyo, Japan}
\affil[2]{Department of Information Sciences, Tokyo University of Science, Tokyo, Japan}
\affil[3]{School of Data Science, Meisei University, Tokyo, Japan}
\affil[4]{RIKEN Center for Brain Science, Saitama, Japan}

\date{}

\begin{document}
\linesparpage{25}
\allowdisplaybreaks[4]
\begin{singlespace}
\maketitle
\end{singlespace}

\section*{Abstract}
In observational studies, the propensity score plays a central role in estimating causal effects of interest. The inverse probability weighting (IPW) estimator is commonly used for this purpose. However, if the propensity score model is misspecified, the IPW estimator may produce biased estimates of causal effects. Previous studies have proposed some robust propensity score estimation procedures. However, these methods require considering parameters that dominate the uncertainty of sampling and treatment allocation. This study proposes a novel Bayesian estimating procedure that necessitates probabilistically deciding the parameter, rather than deterministically. Since the IPW estimator and propensity score estimator can be derived as solutions to certain loss functions, the general Bayesian paradigm, which does not require the considering the full likelihood, can be applied. Therefore, our proposed method only requires the same level of assumptions as ordinary causal inference contexts. The proposed Bayesian method demonstrates equal or superior results compared to some previous methods in simulation experimentss, and is also applied to real data, namely the Whitehall dataset.

\vspace{0.5cm}
\noindent
{\bf Keywords}: covariate balancing, general Bayes, inverse probability weighting, M-estimator, propensity score

\section{Introduction}
Adjusting for covariates (or confounders) is important in estimating the causal effects in observational studies. In the case where all covariates requiring adjustment are observed, it is possible to obtain a consistent estimator for causal effects, commonly known as ``no unmeasured confounding."\cite{He2020} In this context, the propensity score plays a central role in estimating causal effects\cite{Ro1983}. However, because the propensity score is usually unknown, it must be estimated using appropriate procedures, and estimators may be biased if the propensity score model is misspecified\cite{Ka2007}. Therefore, constructing a valid model is crutial for accurate estimation.

Imai and Ratkovic\cite{Im2014} propose the method known as the ``Covariate Balancing Propensity Score" (CBPS), which does not involve modeling the propensity score directly, but, instead estimates the propensity score satisfying a moment restriction. Specifically, the moment condition comprises covariates that are desired to have a balanced distribution between the treatment groups, and the propensity score model is estimated to achieve covariate balance. The CBPS estimation heavily relies on selecting the appropriate moment restriction. Imai and Ratkovic recommend using the CBPS based on second-order moments of covariates to achieve balance. Huang et al.\cite{Hu2022} argue that achieving balance for higher-order moments may be necessary to obtain more accurate estimates of causal effects. More recently, to achieve a more precise covariate balance, several methods have been proposed. For example, kernel balancing methods\cite{Wo2018,Ha2020} aim to improve the balance by using kernel functions. Additionally, a method that satisfies the balancing of the characteristic function of the covariates has been proposed\cite{Hu2020}. However, their methods require exact fulfillment of covariate balancing conditions, implying an equational constraint on a moment condition is necessary. Zubizarreta\cite{Zu2015} argues that achieving an exact balance is challenging because the balance condition depends on the law of large numbers, which is applicable only in infinite samples. In other words, the condition does not account for the uncertainty derived from random sampling and treatment allocation. To address this issue, Zubizarreta proposes the ``stable balancing weights" (SBW) method\cite{Zu2015}, which ensures a covariate balance up to a specified upper bound $\lambda$ that can be determined by researchers. In other words, the covariate balancing condition is subject to the subjective decision of the researcher. It is worth noting that Tan\cite{Ta2020} demonstrates that the propensity score obtained using a logistic regression model that satisfies the same condition as the SBW method can be considered as the solution for a loss function. This result is highly relevant and critical for our proposed procedure.

The SBW-based method depends on moment conditions for covariate balance and the upper bound parameter $\lambda$, which determines the acceptable range of deviation for the covariate balancing condition. As mentioned in the previous paragraph, the upper bound is subjectively determined by the researcher. This study adopts the idea of Bayesian methods to propose a methodology that relaxes the decision process of the parameter. Specifically, we leverage the conclusion that a causal effect estimator (we consider the inverse probability weighting [IPW] estimator) and the SBW-based method can be described as solutions for a loss function. Then, we proceed to specify prior distributions for each model. Additionally, to achieve the mentioned objective, we specify a prior distribution for $\lambda$. To derive the posterior distribution, which plays a central role in Bayesian inference, we employ the general Bayes procedure \cite{Bi2016}, which updates the prior belief with loss functions. 
Using the Bayesian inference procedure, our proposed method can construct credible intervals and perform statistical tests within the Bayesian framework, not only for the causal effects of interest but also for the parameter $\lambda$.

To the best of our knowledge, this study is the first to estimate the IPW estimator from a Bayesian perspective with a covariate balancing condition. The remainder of this paper is organized as follows. In Section 2, we introduces the notations and significant previous works. We specifically demonstrate that the SBW-based propensity score estimation method and the IPW estimator can be explained through certain loss functions. In Section 3, we propose a Bayesian-based estimation procedure using the General Bayes paradigm. Section 4 presents simulation experiments to evaluate the performance of our proposed method against that of some competitors. In Section 5, we apply the methods to real data: Whitehall dataset.

\section{Preliminaries}
\label{sec:pre}
In this section, we introduce covariate balancing methods and their properties, specifically the CBPS\cite{Im2014} and SBW method\cite{Zu2015,Ta2020}. Furthermore, we demonstrate that the IPW estimator, used for estimating causal effects, can be considered as the solution for a loss function. Although this information is well-known, it forms a crucial foundation for the subsequent discussions in this section.

Let $n$ denote the sample size. In the following discussion, we implicitly assume the stable unit treatment value assumption (SUTVA) \cite{Ro1983}. $A_{i}\in\{0,\, 1\}$, $\bld{X}_{i}\in\mathbb{R}^{p}$ and $(Y_{1i},\, Y_{0i})\in \mathbb{R}^{2}$ denote the treatment, a vector of covariates measured prior to treatment, and potential outcomes, respectively. Additionally, from the SUTVA, the observed outcome is $Y_{i}:=A_{i}Y_{1i}+(1-A_{i})Y_{0i}$. This formulation implicitly relies on the (causal) consistency assumption\cite{Ha2020}. Under these settings, we consider that i.i.d.~copies $(A_{i},\bld{X}_{i},Y_{i})$, $i=1,\, 2,\, \dots,\, n$ are obtained. Next, we introduce the Average Treatment Effect (ATE) as the central interest for causal inference, denoted by $\tau:={\rm E}[Y_{1}-Y_{0}]$. Assuming strong ignorability of the treatment assignment, which consists of conditional exchangeability and positivity\cite{Ha2020}, we can estimate the ATE using the propensity score $e(\bld{X}_{i}):={\rm Pr}(A=1\mid\bld{X}_{i})$\cite{Ro1983}.

In observational studies, the propensity score is usually unknown\cite{Im2015}. Therefore, we need to specify the model and derive the predictor of the propensity score for the subsequent analysis. Generally, the propensity score model is assumed to be the logistic regression model. However, model misspecification can lead to crucial bias in the main analysis, that is, when estimating the causal effects. To overcome the problem, Imai and Ratkovic propose the CBPS, which involves estimating the parameter $\bld{\alpha}$ of the propensity score model $e(\bld{X}_{i};\bld{\alpha})$ by satisfying the following moment restriction\cite{Im2014}:
$$
\frac{1}{n}\sum_{i=1}^{n}\left(\frac{A_{i}}{e(\bld{X}_{i};\bld{\alpha})}-\frac{1-A_{i}}{1-e(\bld{X}_{i};\bld{\alpha})}\right)g(\bld{X}_{i})=\bld{0},
$$
where $g(\cdot)$ is a measurable finite dimensional ($L$ dimension) function related to the covariates; for instance (here, $L=2p$),
$$g(\bld{X}_{i})=\left(X_{i1},\dots,X_{ip},X_{i1}^2,\dots,X_{ip}^2\right).$$
Note that when the propensity score model is correctly specified or misspecified locally, the CBPS can derive the consistent estimator for the coefficients $\bld{\alpha}$ \cite{Fa2022}.

In the context of missing data, Zubizarreta\cite{Zu2015} proposes the SBW method. To adapt it to the causal inference context, the SBW-based method can be considered as estimating $\bld{\alpha}$ using the following moment restriction (cf. Tan\cite{Ta2020}):
\begin{align}
\label{ineq1}
\frac{1}{n}\left\vert\sum_{i=1}^{n}\left(\frac{A_{i}}{e(\bld{X}_{i};\bld{\alpha})}-\frac{1-A_{i}}{1-e(\bld{X}_{i};\bld{\alpha})}\right)g_{j}(\bld{X}_{i})\right\vert<\lambda\ \ \ (j=1,2,\dots,L),
\end{align}
where $\lambda$ ($>0$) is a hyper-parameter that determines the upper bound of the covariate balance. In the special case, when $\lambda = 0$, the SBW-based method is obviously consistent with the CBPS. The $\lambda$ can be considered as a trade-off parameter between achieving exact balance for samples and maintaining sampling and treatment allocation randomness\cite{Zu2015}. Zubizarreta demonstrates that the upper bound $\lambda$ influences the bias of the mean of interest for the marginal expectation of outcome \cite{Zu2015}. Theoretically, Wang and Zubizarreta show that this upper bound must converge to $0$ with a certain probability order to ensure consistency \cite{Wa2020}. Furthermore, they discuss a bootstrap sample-based algorithm for determining $\lambda$.

\subsection{Loss Function for the Propensity Score Estimation using Covariate Balancing Procedure}
Tan\cite{Ta2020} derives an important conclusion, stating that the SBW-based method can be considered as the solution of a loss function when the propensity score model is assumed to be the logistic regression model. Specifically, the loss function is described as follows:
\begin{align}
\label{eq1}
\ell_{A}(\bld{\alpha})=\sum_{i=1}^{n}\left[A_{i}\exp\left\{-\bld{\alpha}^{\top}g(\bld{X}_{i})\right\}+(1-A_{i})\bld{\alpha}^{\top}g(\bld{X}_{i})+(1-A_{i})\exp\left\{\bld{\alpha}^{\top}g(\bld{X}_{i})\right\}-A_{i}\bld{\alpha}^{\top}g(\bld{X}_{i})\right]
\end{align}
with the following penalty term:
\begin{align}
\label{pen1}
\lambda\|\bld{\alpha}\|_{1}:=\lambda\sum_{j=1}^{L}\vert\alpha_{j}\vert.
\end{align}
Note that the relationship between (\ref{ineq1}), and (\ref{eq1}) and (\ref{pen1}) is derived from the duality of a constrained optimization problem. For more details, see Appendix \ref{appA}.

\subsection{Loss Function for the IPW Estimator}
After estimating the propensity score, for instance through the loss function (\ref{eq1}), we proceed to estimate the interesting causal effects using methods such as the IPW estimator, which is described as follows:
\begin{align}
\hat{\tau}=\hat{\theta}_{1}-\hat{\theta}_{0}=\frac{\sum_{i=1}^{n}\frac{A_{i}Y_{i}}{e_{i}(\hat{\bld{\alpha}})}}{\sum_{i=1}^{n}\frac{A_{i}}{e_{i}(\hat{\bld{\alpha}})}}-\frac{\sum_{i=1}^{n}\frac{(1-A_{i})Y_{i}}{1-e_{i}(\hat{\bld{\alpha}})}}{\sum_{i=1}^{n}\frac{1-A_{i}}{1-e_{i}(\hat{\bld{\alpha}})}} \label{IPW_est},
\end{align}
where $\hat{\bld{\alpha}}$ is an estimator including the solution of (\ref{eq1}). Through the well-known result that the IPW estimator can be described as an M-estimator, the $\hat{\theta}_{k}\, $s are obtained as the solution to the loss functions $\ell_{Y_{1}}(\theta_{1},\hat{\bld{\alpha}})$ and $\ell_{Y_{0}}(\theta_{0},\hat{\bld{\alpha}})$, where
\begin{align}
\ell_{Y_{1}}(\theta_{1},\bld{\alpha})&=\sum_{i=1}^{n}\frac{A_{i}}{e_{i}(\bld{\alpha})}\left(Y_{i}-\theta_{1}\right)^2, \label{eq2}
\end{align}
and
\begin{align}
\ell_{Y_{0}}(\theta_{0},\bld{\alpha})&=\sum_{i=1}^{n}\frac{1-A_{i}}{1-e_{i}(\bld{\alpha})}\left(Y_{i}-\theta_{0}\right)^2 \label{eq3}.
\end{align}
From the above discussions, the parameters of interest, namely $\bld{\alpha}$, $\theta_{1}$, and $\theta_{0}$, can be obtained as solutions to the corresponding loss functions (\ref{eq1}), (\ref{eq2}), and (\ref{eq3}), respectively. In the following sections, we construct a Bayesian inference procedure by using the loss functions.

\section{Bayesian Estimating Procedures}
In this section, we propose a Bayesian inference procedure for the interesting causal effect $\tau$ based on the general Bayesian framework\cite{Bi2016}. 
Specifically, we introduce prior distributions and sampling procedures for each parameter's posterior distribution.
\subsection{Brief Introduction of General Bayes}
First of all, we briefly introduce the general Bayesian updating proposed by Bissiri et al.\cite{Bi2016} and connect it to the discussion in this manuscript. In this framework, to incorporate data information into prior beliefs, likelihood functions are not necessarily required; instead, a broader class of loss functions is accepted. 
The posterior, referred to as \textit{the generalized posterior}, is defined as 
\begin{equation*}
    p_{G}(\theta | \bm{x}) \propto p(\theta) \exp\{ - \omega \ell(\theta, \bm{x}) \},
\end{equation*}
where $\bm{x} = (x_1,\ldots,x_n)$ is an observed sample, $p(\theta)$ is the prior distribution of the parameter $\theta$, and $\ell(\theta, \bm{x})$ is a loss function satisfying the \textit{additivity}, i.e., $\ell(\theta, \bm{x}) = \sum_{i=1}^n \ell(\theta,x_i)$.
The learning rate $\omega$, also referred to as the loss scale or calibration weight, is a non-negative scalar that regulates the learning for the parameter $\theta$ from the observed data.
If $\omega$ is too large, the posterior distribution will concentrate excessively, inflating the importance of the data information about $\theta$. 
Conversely, if $\omega$ is too small, the posterior distribution will underweight the information from the data relative to the prior.
Note that Bayes' rule is recovered as a special case when $\omega=1$ and $\ell(\theta, \bm{x})$ is the negative log-likelihood function.
Although this framework was suggested by some earlier works\cite{Yi2009}, Bissiri et al. provided a proof of the coherence and rationality of Bayesian inference, even when using general loss functions\cite{Bi2016}. Therefore, the use of the loss functions (\ref{eq1}), (\ref{eq2}), and (\ref{eq3}) is justified within the context of Bayesian inference.

\subsection{Posterior of General Bayesian Estimation}
To construct the posterior inference procedure for $\theta_1$ and $\theta_0$, we formulate the generalized posterior
\begin{equation}
\label{eq:gb1}
p_G(\theta_1, \theta_0, \bm{\alpha}, \lambda \mid \bm{Y},\bm{X},\bm{A}) = p_{G_{\theta_1}}(\theta_1 \mid \bm{Y},\bm{X},\bm{A},\bm{\alpha}) p_{G_{\theta_0}}(\theta_0 \mid \bm{Y},\bm{X},\bm{A},\bm{\alpha}) p_{G_{\bm{\alpha}}}(\bm{\alpha}, \lambda \mid \bm{X},\bm{A}), 
\end{equation}
where
\begin{equation}
\label{eq:gp_theta1}
p_{G_{\theta_1}}(\theta_1 \mid \bm{Y},\bm{X},\bm{A},\bm{\alpha}) = C_1 p(\theta_1 \mid \bm{\alpha}) \exp\{ -\omega \ell_{Y_1}(\theta_1, \bm{\alpha}) \},
\end{equation}
\begin{equation}
\label{eq:gp_theta0}
p_{G_{\theta_0}}(\theta_0 \mid \bm{Y},\bm{X},\bm{A},\bm{\alpha}) = C_0 p(\theta_0 \mid \bm{\alpha}) \exp\{ -\omega \ell_{Y_0}(\theta_0, \bm{\alpha}) \},
\end{equation}
and
\begin{equation}
\label{eq:gp_alpha}
p_{G_{\bm{\alpha}}}(\bm{\alpha}, \lambda \mid \bm{X},\bm{A}) = C_a p(\bm{\alpha}\mid\lambda) p(\lambda) \exp\{ -\omega \ell_A(\bm{\alpha}) \}.
\end{equation}
Normalized constants $C_1$, $C_0$, and $C_a$ are independent of $\theta_1$, $\theta_0$, and $\bm{\alpha}$ and $\lambda$, respectively.
The priors of $\theta_1$, $\theta_0$, $\bm{\alpha}$, and $\lambda$ are denoted by $p(\theta_1 \mid \bm{\alpha})$, $p(\theta_0 \mid \bm{\alpha})$, and $p(\bm{\alpha}\mid\lambda)$, and $p(\lambda)$, respectively.
The $\omega$ denotes the learning rate, and the method to determine its value is described later.
Note that assuming a uniform prior for $\theta_1$ and $\theta_0$ and a Laplace (or double exponential) prior with a location parameter of 0 and fixed scale parameter of $\lambda$ for $\bm{\alpha}$, the maximum a posteriori (MAP) estimator of the general posterior \eqref{eq:gb1}, where $\omega=1$ is equivalent to the estimator obtained by the estimation procedure through loss functions \eqref{eq1}, \eqref{eq2}, and \eqref{eq3} described in Section \ref{sec:pre}. 

While the general posteriors \eqref{eq:gp_theta1} and \eqref{eq:gp_theta0} depend on $\bm{\alpha}$ and $\lambda$, the general posterior \eqref{eq:gp_alpha} is independent of $\theta_1$ and $\theta_0$.
Hence, we can obtain the marginal posterior of $\theta_1$ and $\theta_0$ as follows.

\vspace{0.5cm}
\noindent
{\bf {\large Sampling schemes}}\\
\begin{itemize}
\item[Step.1] Obtain posterior draws $\bm{\alpha}^{(r)}$ and $\lambda^{(r)}$ from the general posterior \eqref{eq:gp_alpha} for each iterations $r=1,2,\ldots,R$.
\item[Step.2] Obtain posterior draws $\theta_k^{(r)}$ from the general posteriors \eqref{eq:gp_theta1} and \eqref{eq:gp_theta0} of $\theta_k$ for $k=0,1$.
\end{itemize}
\vspace{0.5cm}

The sampling of $\bm{\alpha}$ and $\lambda$ in Step.1 can be implemented using the loss likelihood bootstrap\cite{Ne2021} or probabilistic programming language \texttt{Stan}\cite{Ca2017}.
In the implementation, we use the Gamma prior for $\lambda$.

For Step.2, $\theta_k$ can be drawn from the normal distribution. 
Suppose that the prior of $\theta_k$ is the Normal distribution with the mean $\mu_k$ and precision $\tau_k$, denoted as $N(\mu_k,\tau_k)$. The general posterior of $\theta_k$ is expressed as
\begin{align*}
p(\theta_k \mid \bm{Y}, \bm{X}, \bm{A}, \bm{\alpha}) &\propto \exp\left\{ -\frac{\tau_k}{2} (\theta_k-\mu_k)^2 \right\} \exp\left\{ -\frac{\omega}{2} \sum_{i=1}^n s_{ki} (\theta_k-Y_i)^2 \right\} \\
&\propto \exp\left\{ -\frac{\Tilde{\tau}_k}{2} (\theta_k - \Tilde{\mu}_k)^2 \ \right\}
\end{align*}
where
\begin{equation*}
s_{ki} = \frac{2 A_i^k(1-A_i)^{1-k}}{e_i(\bm{\alpha})^k \{1-e_i(\bm{\alpha})\}^{1-k} }, ~~ \Tilde{\mu}_k = \Tilde{\tau}_k^{-1} \left(\tau_k\mu_k + \omega\sum_{i=1}^n s_{ki}Y_i \right) ~~ \mbox{and} ~~ \Tilde{\tau}_k = \tau_k + \omega\sum_{i=1}^n s_{ki}.
\end{equation*}
Therefore, in Step.2, we can easily obtain the posterior draws $\theta_k^{(r)}$ from $N(\Tilde{\mu}_k^{(r)}, \Tilde{\tau}_k^{(r)})$, where
\begin{equation*}
\Tilde{\mu}_k^{(r)} = \Tilde{\tau}_k^{(r)} \left(\tau_k\mu_k + \omega\sum_{i=1}^n s_{ki}^{(r)}Y_i \right) ~~ \mbox{and} ~~ \Tilde{\tau}_k^{(r)} = \tau_k + \omega\sum_{i=1}^n s_{ki}^{(r)}, ~~ s_{ki}^{(r)} = \frac{2 A_i^k(1-A_i)^{1-k}}{e_i(\bm{\alpha}^{(r)})^k \{1-e_i(\bm{\alpha}^{(r)})\}^{1-k} }.
\end{equation*}
The derivation of the posterior distribution is provided in Appendix \ref{appA}.

In the proposed sampling schemes, the posterior sampling scheme is conducted in two stages. First, parameters related to the posterior of the propensity score model are sampled, followed by parameters for the IPW estimator. This approach corresponds to the ``two-step inference" known in the Bayesian context \cite{Zi2014,Li2020,St2023}. Another Bayesian procedure applied in the context of causal inference is called ``joint modeling"; however, these methods have both strengths and weaknesses \cite{Li2023}. In this manuscript, we select the former procedure.

\begin{remark}{\it About Tuning the Learning Rate $\omega$.}\\
Since the learning rate $\omega$ affects the inaccuracy of uncertainty quantification in the generalized Bayesian posterior, a number of studies exist on its tuning\cite{grunwald2017inconsistency, holmes2017assigning, lyddon2019general, syring2019calibrating}.
Most of these studies involve ad hoc fine-tuning in a frequentist framework and are computationally complex and expensive.
In contrast to these, we consider a simple and effective learning rate selection based on the posterior covariance information criterion (PCIC)\cite{Ib2022,Ib2023}, which can be easily computed using the posterior draws from the generalized Bayesian posterior \eqref{eq:gb1}.
The PCIC is a natural generalization of the widely applicable information criterion (WAIC)\cite{watanabe2010equations} and has the property of being an asymptotic equivalence to the Bayesian leave-one-out cross-validation\cite{gelfand1994bayesian}.

The PCIC in our generalized Bayesian procedure is computed as follows:
\begin{equation*}
    \begin{split}
        {\rm PCIC} = {} & \frac{1}{n} \sum_{i=1}^n \frac{1}{R} \sum_{r=1}^R \nu(Y_i, A_{i}, \bm{X}_i \mid \theta_1^{(r)}, \theta_0^{(r)}, \bm{\alpha}^{(r)}) \\
        {} & - \left[ \frac{1}{n} \sum_{i=1}^n \frac{1}{R} \sum_{r=1}^R \left\{ \nu(Y_i, A_{i}, \bm{X}_i \mid \theta_1^{(r)}, \theta_0^{(r)}, \bm{\alpha}^{(r)}) s(Y_i, A_{i}, \bm{X}_i \mid \theta_1^{(r)}, \theta_0^{(r)}, \bm{\alpha}^{(r)}) \right\} \right. \\
        {} & \quad \quad - \left. \frac{1}{n} \sum_{i=1}^n \left\{ \frac{1}{R} \sum_{r=1}^R \nu(Y_i, A_{i}, \bm{X}_i \mid \theta_1^{(r)}, \theta_0^{(r)}, \bm{\alpha}^{(r)}) \right\} \left\{ \frac{1}{R} \sum_{r=1}^R s(Y_i, A_{i}, \bm{X}_i \mid \theta_1^{(r)}, \theta_0^{(r)}, \bm{\alpha}^{(r)}) \right\} \right]
    \end{split}
\end{equation*}
where $s(\cdot)$ is a score function that is the negative sum of equations \eqref{eq1}, \eqref{eq2} and \eqref{eq3}, while $\nu(\cdot)$ is an arbitrary loss function for a future observation and the parameter vector $\theta_1$, $\theta_0$ and $\bm{\alpha}$.
Here, we set $\nu(\cdot) = s(\cdot)$ since this implies a learning rate $\omega$ that reduces the prediction risk with respect to the loss function \eqref{eq1}, \eqref{eq2}, and \eqref{eq3}.

The R program available to the GitHub repository for using our proposed method outputs the PCIC calculated for several learning rates (0.2, 0.5, 1.0, and 1.5 by default), and we only need the analysis result with the learning rate that gives the minimum PCIC.
\end{remark}

By applying the PCIC, the predicted risks associated with the loss functions \eqref{eq1}, \eqref{eq2}, and \eqref{eq3} is minimized. Focusing on the loss functions related to the causal effect (\eqref{eq2} and \eqref{eq3}), this process is similar to the one discussed in Saarela et al.\cite{Sa2015}. Specifically, they consider a predicted (weighted) utility function for the target parameters; in their procedure, the utility is defined as a likelihood function related to the outcome. For instance, for the treatment group, the loss function becomes:
$$
U(\theta_{1})=\sum_{i=1}^{n}\frac{A_{i}}{e_{i}(\hat{\bld{\alpha}})}\log\left[\frac{1}{\sqrt{2\pi\sigma^2}}\exp\left\{-\frac{1}{2\sigma^2}\left(Y_{i}-\theta_{1}\right)^2\right\}\right]\propto\sum_{i=1}^{n}\frac{A_{i}}{e_{i}(\hat{\bld{\alpha}})}\left(Y_{i}-\theta_{1}\right)^2,
$$
where $\sigma^2$ is the variance parameter of the outcome. It can be viewed that their approach shares the same spirit as our PCIC-based procedure.

\section{Simulation Experiments}
In this section, we compare the performance of our proposed method with that of competing methods using simulation datasets. The setting is based on Leacy and Stuart\cite{Le2014}, Setodji et al\cite{Se2017}, and Orihara et al.\cite{Or2022}. 
The iteration number for all simulation examples was $10,000$. Some materials are appeared in Appendix \ref{appB}.

\subsection{Data-generating Mechanism}
First, we describe the data-generating mechanism used in the simulations. Assuming that there were ten covariates, $\bld{X}_{i}=(X_{i1},\dots,X_{i10})$. $X_{i1}$, $X_{i3}$, $X_{i6}$, and $X_{i9}$ were generated from $Ber(0.5)$ and $X_{i7}$ and $X_{i10}$ were generated from the standard normal distribution. $X_{i2}$ and $X_{i4}$ were generated from $N(x_{i6},0.1^2)$ and $N(x_{i9},0.1^2)$, respectively. $X_{i5}$ and $X_{i8}$ were generated from $Ber(expit\left\{0.4(2x_{i1}-1)\right\})$ and $Ber(expit\left\{0.4(2x_{i3}-1)\right\})$, respectively. In this setting, the correlations between the following combinations of covariates were approximately:
$$
Corr(X_{1},X_{5})=Corr(X_{3},X_{8})=0.2,\ Corr(X_{2},X_{6})=Corr(X_{4},X_{9})=0.9,
$$
and the correlations of the other combinations were approximately $0$.

Next, we introduce the assignment mechanism of the treatment value $A_{i}$; that is, the true propensity score $e(\bld{X}_{i})={\rm Pr}\left(A=1\mid\bld{X}_{i}\right)=expit\left\{h_{A}\left(\bld{X}_{i}\right)\right\}$. Here, we considered six scenarios by changing the the linear predictor $h_{A}\left(\bld{X}_{i}\right)$; specifically, {\bf (a)}:\ ``Small complexity and good overlap,'' {\bf (b)}:\ ``Small complexity and poor overlap,'' {\bf (c)}:\ ``Moderate complexity and good overlap,'' {\bf (d)}:\ ``Moderate complexity and poor overlap,'' {\bf (e)}:\ ``Large complexity and good overlap,'' and {\bf (f)}:\ ``Large complexity and poor overlap.'' For more details, see Appendix \ref{appB} and Orihara et al.\cite{Or2022}.

Finally, we introduce the model for the outcome $Y_{i}$. This simulation considered the binary outcome and logistic regression model ${\rm Pr}\left(Y=1\mid\bld{X}_{i}\right)=expit\left\{h_{Y}\left(\bld{X}_{i}\right)\right\}$ where
\begin{align}
\label{t_out}
h_{Y}^{\top}\left(\bld{X}_{i}\right)&=\left(1,A_{i},A_{i}\times X_{i2},A_{i}\times X_{i4},X_{i1},\dots,X_{i4},X_{i8},\dots,X_{i10},X_{i2}^2,\right.\nonumber\\
&\hspace{0.5cm}\left.X_{i1}\times X_{i3},X_{i2}\times X_{i4},X_{i4}\times X_{i8},X_{i8}\times X_{i9}\right)\bld{\xi},
\end{align}
$$
\bld{\xi}=\left(-2,0.2,1,1,0.3,-0.36,-0.73,-0.2,0.71,-0.19,0.26,-0.36,0.15,-0.252,-0.1,0.355\right)^{\top}.
$$
Under these settings, the true causal effects (i.e., ATE) is approximately $0.152$ for all propensity score situations {\bf (a)}--{\bf (f)}.

\subsection{Estimating Methods}
We considered four methods for estimating the propensity score: ordinary logistic regression, covariate balancing propensity score (CBPS\cite{Im2014}), regularized calibration method (RCAL\cite{Ta2020}), a Bayesian method using a Logistic regression-based Propensity Score (BLPS) proposed by Saarela et al. \cite{Sa2015}, and the proposed Bayesian method explained in Section 3.

To estimate the propensity score, we consider two situations:\ ``Confounders" and ``Predictors." For the former situation, we use only the confounders to estimate the propensity score; $g(\bld{X}_{i})=\left(X_{i1},X_{i2},X_{i3},X_{i4}\right)$. Thus, we can at least adjust for confounder bias. In the latter situation, we use the predictors of the true propensity score; $g(\bld{X}_{i})=\left(X_{i1},X_{i2},\dots,X_{i7}\right)$. Thus, there is a chance to estimate an unbiased propensity score compared to the ``Confounders" situations. However, note that adjusting for variables that are only related to the exposure variable ($X_{i5}$, $X_{i6}$, and $X_{i7}$) can yield an increase in the variance of causal effect estimators\cite{Br2006}.

As mentioned in Section 2, the CBPS is a robust parametric propensity score estimation method that considers model misspecification. In this simulation experiment, we only considered the first-order moment aligning with previous research\cite{Se2017}. Note that other methods also use the same moment condition, and the covariates are standardized.

For the RCAL method proposed in Tan\cite{Ta2020}, it is necessary to decide on the tuning parameter $\lambda$, as described in Section 2 (see (\ref{ineq1})). Specifically, to determine this parameter, we consider two approaches: making a deterministic decision or using Cross-Validation (CV). In the former approach, we prepare some values of $\lambda$ in advance. In the latter approach, CV is conducted with the loss function related to the propensity score (\ref{eq1}). To implement the RCAL, we employ the \texttt{glm.regu} and \texttt{ate.ipw} functions from the \texttt{RCAL} package in \texttt{R}.

For the BLPS method proposed by Saarela et al. \cite{Sa2015}, the Bayesian bootstrap is used to estimate the ATE. While their method can be applied to sequential treatment settings, we used it only for a point treatment setting.

The proposed method, called Bayesian RCAL (BRCAL), is based on an estimation procedure that includes determining the learning rate $\omega$ using PCIC, as explained in Section 3. Prior distributions of parameters are $\lambda\sim Gam(0.01, 0.1)$, $\bld{\alpha}\sim Double Exponential(0,1/\lambda)$, and $\theta_{k}\sim N(0,100^2)$ ($k=0,1$). The specific sampling schemes are described in Section 3.2. The program files for implementing BLPS and our proposed method are stored on the author's GitHub page.

\subsection{Performance Metrics}
We evaluated the various methods based on bias, root mean squared error (RMSE), coverage probability (CP), average length of Confidence (Credible) Interval (AvL), bias and RMSE ratio (BR/RR) compared to the proposed method, and boxplot of estimated ATE from 10,000 iterations. The bias and RMSE were calculated as $bias=\bar{\hat{\tau}}-\tau_{0}$, $RMSE=\sqrt{\frac{1}{10000}\sum_{k=1}^{10000}\left(\hat{\tau}_{k}-\tau_{0}\right)^2}$, where $\bar{\hat{\tau}}=\frac{1}{10000}\sum_{k=1}^{10000}\hat{\tau}_{k}$, $\hat{\tau}_{k}$ is the estimate of each estimator and iteration, and $\tau_{0}$ ($=0.152$) is the true value of the ATE. The CP refers to the proportion of cases where the confidence interval or credible interval includes $\tau_{0}$. The AvL was calculated as the average length of the Confidence (Credible) Interval over 10,000--time iterations. Bias and RMSE ratio (BR/RR) are calculated as the ratio compared to the bias/RMSE of our proposed method. 

To evaluate the performance metrics of the BLPS and BRCAL, the posterior mean was used to assess the bias, RMSE, and BR/RR. For CP and AvL, the posterior distribution was used.

\subsection{Simulation Results}
The results are summarized in Tables \ref{tab1}--\ref{tab3} and Figure \ref{fig1}, with additional results provided in Appendix \ref{appB}. In the main manuscript, we only discuss the results under ``Confounders" situations.

In small sample situations, all methods exhibit small bias in scenarios {\bf (a)}-–{\bf (d)}. However, in scenarios {\bf (e)} and {\bf (f)}, where the true propensity score model is complex, all methods---especially RCAL---yield large bias. This bias improves in moderate and large sample situations. The CBPS and BRCAL show similar performance, and notably, both methods exhibit the smallest bias in scenarios {\bf (e)} and {\bf (f)}.

Across all scenarios {\bf (a)}--{\bf (f)}, the RMSE values are nearly identical for all methods. To some extent, the RCAL performs slightly better in small sample situations.

The 95\% coverage probability (CP) of the BRCAL nearly achieves the nominal level and is higher than that of other methods in scenarios {\bf (a)}--{\bf (d)}. In scenario {\bf (e)}, the BRCAL slightly underestimates the 95\% CI, but similar tendencies are observed in other methods. In small samples, all methods except the BRCAL and BLPS tend to underestimate the 95\% CI in many scenarios. While the BLPS achieves the nominal level, it is overly conservative.

The BRCAL, our proposed Bayesian method, can be regarded as the Bayesian counterpart of the RCAL. However, its performance closely resembles that of the CBPS, particularly in large samples. This similarity may be explained by covariate balancing. For more details, see Appendix B.3.

To summarize, our proposed method demonstrates superior or at least comparable performance to other methods. Notably, it provides more accurate causal effect estimates and reliable inference, particularly in small samples. Covariate balancing diagnostics using standardized mean differences (SMDs) and sensitivity analysis results for the prior distribution of $\lambda$ are presented in Appendix B. Based on these analyses, we conclude that our proposed method achieves effective covariate balancing (similar to the CBPS) and remains robust to the choice of the prior distribution for $\lambda$.

\hspace{5cm}[Tables \ref{tab1}--\ref{tab3} and Figure \ref{fig1} here.]

\section{Illustrative Real Data Example}
In this section, we provide an illustrative example of a real data analysis conducted on the cohort study for male civil servants in London (Whitehall dataset)\cite{Ro1999}. Specifically, we examined the causal effects of smoking on all-cause mortality. To estimate these effects, we considered three confounders:\ age, plasma cholesterol concentration, and Civil Service grade. To account for the binary exposure, we coded the original data with $A_{i}=0$ for non-smokers, and $A_{i}=1$ otherwise. The study included 17,260 subjects, of which 7,157 (41.5\%) were smokers. There were also 1,670 deaths (9.7\%). For more details, see Appendix \ref{appC}.

The summary results of the causal effects for each method are as follows: The crude mean difference is $6.86$ $(5.97, 7.75)$ (point estimate (95\% CI)), the CBPS result is $5.62$ $(5.58, 5.68)$, and the RCAL result is $5.71$ $(4.81, 6.62)$. The result of our proposed method is $5.65$ $(4.19, 7.12)$ (posterior mean (95\% CI)). Note that the specific sampling schemes are described in Section 3.2, and prior distributions of parameters are $\lambda\sim Gam(0.01, 0.1)$, $\bld{\alpha}\sim Double Exponential(0,1/\lambda)$, and $\theta_{k}\sim N(0,100^2)$ ($k=0,1$). The point estimates are similar for the CBPS, RCAL, and BRCAL. The 95\% CI for the BRCAL is the widest among the three methods. Given that there are 17,260 subjects, the CI is considered too wide. However, the CI is expected to provide conservative results, based on the findings from the simulation experiments.

\section{Discussions and Conclusions}
In this manuscript, we introduce a novel Bayesian-based IPW estimator that incorporates a covariate balancing condition. Using the general Bayes framework, which is gaining interest in Bayesian analysis, our proposed method does not require consideration of the full likelihood. Accordingly, it only necessitates minimal moment conditions, as is common in causal inference contexts. Our approach demonstrates reasonable performance, especially in small sample scenarios, as shown by various simulations and real data analyses.

To the best of our knowledge, this is the first method that applies the general Bayes framework to causal inference with a covariate balancing condition. As described in Section 3, the learning rate $\omega$ is determined by PCIC. This method proves effective, as evidenced by the 95\% CI aligning well with the nominal probability. However, relying on PCIC is an ad-hoc strategy lacking theoretical justification. In future work, we aim to develop a consensus method that is theoretically grounded.

In our proposed framework, as mentioned in section 3.2, the posterior sampling scheme is conducted in two stages which allows us to estimate causal effects as in standard causal inference contexts \cite{Im2015}. The inference is justified from a Bayesian perspective \cite{Li2020}, provided the second step is likelihood-based. Bissiri et al. \cite{Bi2016} note that the general Bayesian procedure ensures valid Bayesian inference by meeting coherence and rationality \cite{Be2013}. Future work should explore further justification when applying the general Bayesian framework for the second step.

While simpler approaches may substitute the second step with a direct plug-in of the propensity score posterior into the IPW estimator (\ref{IPW_est}), our method is preferable. It marginalizes over the propensity score model\cite{Li2020}, potentially mitigating misspecification issues\cite{Ka2007}. Furthermore, a doubly robust estimator --- valid if either the propensity score or outcome model is correctly specified --- has been proposed within the Bayesian context\cite{Sa2016}, offering a competitive alternative to our method.

For future work, applying our approach to sequential treatment settings is crucial for estimating the ATE in situations with variable doses. While some methods incorporating covariate balancing conditions have been proposed \cite{Im2015_CBPS,Av2021}, a loss function akin to that described by Tan \cite{Ta2020} has yet to be developed. Going forward, deriving such a loss function and considering a Bayesian estimation strategy following the same approach as presented in this manuscript will be essential. In sequential treatment settings, concerns about missing data, particularly in scenarios with extended follow-up periods, arise. In Appendix \ref{appD}, we consider the situations where one covariate is missing in point treatment scenarios. However, these methods cannot be directly applied to sequential settings. Xu et al. \cite{Xu2016} introduce the Bayesian Additive Regression Tree method for imputing missing data as a Bayesian approach. This method, which is potentially adaptable to our proposed and future methodologies, could enable the construction of a comprehensive Bayesian method that effectively handles missing covariate data.

\vspace{0.5cm}
\noindent
{\bf Author contributions:} All authors have accepted responsibility for the entire content of this paper and approved its submission.

\vspace{0.5cm}
\noindent
{\bf Acknowledgements:} We would like to express our gratitude to the editors and reviewers for their useful comments, which have contributed greatly towards improving the manuscript. We would also like to thank Editage (www.editage.com) for the English language editing. All the authors accept responsibility for the content of this manuscript and approve its submission. 

\vspace{0.5cm}
\noindent
{\bf Declaration of interest:}
The authors declare they have no conflicts of interest.

\vspace{0.5cm}
\noindent
{\bf Data availability statement:}
Data sharing is not applicable to this article as no new data were created or analyzed in this study. The simulation analysis programs are available at the following URL:
\begin{itemize}
\item \url{https://github.com/t-momozaki/BRCAL}
\end{itemize}

\bibliography{bibfile}

\newpage
\begin{landscape}
\begin{table}[htbp]
\begin{center}
\caption{Summary of estimated causal effects under the scenario {\bf (a)} and {\bf (b)} (``Confounder" situation):\ Bias, root mean squared error (RMSE), 95\% coverage probability (95\%CP), average length of CI (AvL), and bias and RMSE ratio (BR/RR) compared to the bias/RMSE of our proposed method of the IPW estimator in 10,000 iterations were summarized for each scenario (in ``Scenario" column) and method (in the ``Method'' column). RCAL:\ CV was used for the CV described in Section 4.2. BRCAL:\ PCIC was used for the PCIC to select the learning rate $\omega$ from candidates: $(0.2, 0.5, 1.0, 1.5)$ (see Section 4.2).}
\scalebox{0.63}{
\begin{tabular}{ccc|cccccc|cccccc|cccccc}\hline
{\bf Scenario}&\multicolumn{2}{c|}{\bf Method}&\begin{tabular}{c}{\bf Bias}\\$(\times 10^{-2})$\end{tabular}&\begin{tabular}{c}{\bf RMSE}\\$(\times 10^{-2})$\end{tabular}&{\bf 95\%CP}&{\bf AvL}&{\bf BR}&{\bf RR}&\begin{tabular}{c}{\bf Bias}\\$(\times 10^{-2})$\end{tabular}&\begin{tabular}{c}{\bf RMSE}\\$(\times 10^{-2})$\end{tabular}&{\bf 95\%CP}&{\bf AvL}&{\bf BR}&{\bf RR}&\begin{tabular}{c}{\bf Bias}\\$(\times 10^{-2})$\end{tabular}&\begin{tabular}{c}{\bf RMSE}\\$(\times 10^{-2})$\end{tabular}&{\bf 95\%CP}&{\bf AvL}&{\bf BR}&{\bf RR}\\
\hline
(a) & \multicolumn{2}{c|}{Logit} & 0.722 & 5.671 & 0.940 & 0.215 & 0.952 & 1.006 & 0.289 & 3.493 & 0.948 & 0.136 & 1.116 & 1.002 & 0.252 & 2.462 & 0.950 & 0.096 & 1.058 & 1.001 \\ \cline{2-21} 
 & \multicolumn{2}{c|}{CBPS} & 0.702 & 5.670 & 0.939 & 0.214 & 0.925 & 1.006 & 0.305 & 3.493 & 0.948 & 0.136 & 1.175 & 1.002 & 0.259 & 2.462 & 0.950 & 0.096 & 1.087 & 1.001 \\ \cline{2-21}
 & RCAL & CV & 0.999 & 5.510 & 0.940 & 0.208 & 1.317 & 0.977 & 0.148 & 3.471 & 0.941 & 0.132 & 0.571 & 0.996 & 0.188 & 2.456 & 0.941 & 0.093 & 0.789 & 0.998 \\ \cline{3-21}
 &  & $\lambda=0.01$ & 0.971 & 5.577 & 0.937 & 0.209 & -- & -- & 0.022 & 3.434 & 0.944 & 0.132 & -- & -- & 0.076 & 2.418 & 0.945 & 0.093 & -- & -- \\ \cline{3-21}
 &  & $\lambda=0.05$ & 1.204 & 5.305 & 0.949 & 0.208 & -- & -- & 0.767 & 3.298 & 0.951 & 0.130 & -- & -- & 0.927 & 2.322 & 0.953 & 0.092 & -- & -- \\ \cline{3-21}
 &  & $\lambda=0.10$ & 0.880 & 5.234 & 0.950 & 0.206 & -- & -- & 0.300 & 3.280 & 0.950 & 0.130 & -- & -- & 0.282 & 2.313 & 0.953 & 0.092 & -- & -- \\ \cline{2-21}
 & \multicolumn{2}{c|}{BLPS} & 0.716 & 5.713 & 0.983 & 0.278 & -- & -- & 0.299 & 3.501 & 0.988 & 0.176 & -- & -- & 0.252 & 2.465 & 0.988 & 0.124 & -- & -- \\ \cline{2-21}
 & BRCAL & PCIC & 0.759 & 5.639 & 0.959 & 0.258 & Ref. & Ref. & 0.259 & 3.486 & 0.960 & 0.147 & Ref. & Ref. & 0.238 & 2.460 & 0.962 & 0.104 & Ref. & Ref. \\ \cline{2-21}
 &  & $\omega=0.2$ & 0.834 & 5.498 & 1.000 & 0.619 & -- & -- & 0.112 & 3.446 & 1.000 & 0.392 & -- & -- & 0.110 & 2.447 & 1.000 & 0.277 & -- & -- \\ \cline{3-21}
 &  & $\omega=0.5$ & 0.798 & 5.590 & 1.000 & 0.392 & -- & -- & 0.205 & 3.474 & 0.999 & 0.248 & -- & -- & 0.205 & 2.456 & 0.999 & 0.175 & -- & -- \\ \cline{3-21}
 &  & $\omega=1.0$ & 0.769 & 5.628 & 0.985 & 0.278 & -- & -- & 0.255 & 3.482 & 0.989 & 0.175 & -- & -- & 0.228 & 2.459 & 0.988 & 0.124 & -- & -- \\ \cline{3-21}
 &  & $\omega=1.5$ & 0.743 & 5.643 & 0.955 & 0.227 & -- & -- & 0.260 & 3.486 & 0.957 & 0.143 & -- & -- & 0.242 & 2.460 & 0.959 & 0.101 & -- & -- \\ 
   \hline

(b) & \multicolumn{2}{c|}{Logit} & 0.200 & 6.216 & 0.938 & 0.232 & 1.235 & 1.007 & 0.151 & 3.837 & 0.945 & 0.149 & 0.869 & 1.001 & 0.355 & 2.731 & 0.946 & 0.105 & 0.897 & 1.000 \\ \cline{2-21}
 & \multicolumn{2}{c|}{CBPS} & 0.128 & 6.213 & 0.936 & 0.231 & 0.790 & 1.006 & 0.190 & 3.842 & 0.943 & 0.148 & 1.090 & 1.002 & 0.404 & 2.736 & 0.946 & 0.105 & 1.021 & 1.001 \\ \cline{2-21}
 & RCAL & CV & 0.651 & 5.869 & 0.926 & 0.208 & 4.018 & 0.951 & 0.012 & 3.775 & 0.917 & 0.132 & 0.068 & 0.984 & 0.301 & 2.710 & 0.915 & 0.093 & 0.760 & 0.992 \\ \cline{3-21}
 &  & $\lambda=0.01$ & 0.645 & 6.086 & 0.914 & 0.208 & -- & -- & 0.048 & 3.717 & 0.922 & 0.132 & -- & -- & 0.222 & 2.636 & 0.922 & 0.093 & -- & -- \\ \cline{3-21}
 &  & $\lambda=0.05$ & 0.893 & 5.388 & 0.948 & 0.208 & -- & -- & 1.002 & 3.384 & 0.948 & 0.131 & -- & -- & 0.942 & 2.397 & 0.945 & 0.093 & -- & -- \\ \cline{3-21}
 &  & $\lambda=0.10$ & 0.710 & 5.130 & 0.953 & 0.205 & -- & -- & 0.743 & 3.257 & 0.955 & 0.129 & -- & -- & 0.631 & 2.304 & 0.950 & 0.091 & -- & -- \\ \cline{2-21}
 & \multicolumn{2}{c|}{BLPS} & 0.236 & 6.358 & 0.974 & 0.280 & -- & -- & 0.151 & 3.866 & 0.979 & 0.176 & -- & -- & 0.355 & 2.742 & 0.977 & 0.125 & -- & -- \\ \cline{2-21}
 & BRCAL & PCIC & 0.162 & 6.173 & 0.937 & 0.228 & Ref. & Ref. & 0.174 & 3.834 & 0.942 & 0.146 & Ref. & Ref. & 0.396 & 2.732 & 0.941 & 0.105 & Ref. & Ref. \\ \cline{3-21}
 &  & $\omega=0.2$ & 0.299 & 5.940 & 1.000 & 0.621 & -- & -- & 0.057 & 3.777 & 1.000 & 0.392 & -- & -- & 0.332 & 2.713 & 1.000 & 0.277 & -- & -- \\ \cline{3-21}
 &  & $\omega=0.5$ & 0.231 & 6.098 & 0.999 & 0.393 & -- & -- & 0.140 & 3.815 & 0.999 & 0.248 & -- & -- & 0.378 & 2.726 & 0.999 & 0.175 & -- & -- \\ \cline{3-21}
 &  & $\omega=1.0$ & 0.179 & 6.156 & 0.976 & 0.278 & -- & -- & 0.165 & 3.829 & 0.979 & 0.175 & -- & -- & 0.388 & 2.731 & 0.976 & 0.124 & -- & -- \\ \cline{3-21}
 &  & $\omega=1.5$ & 0.161 & 6.174 & 0.935 & 0.227 & -- & -- & 0.175 & 3.835 & 0.937 & 0.143 & -- & -- & 0.394 & 2.732 & 0.935 & 0.101 & -- & -- \\ \hline
\end{tabular}
}
\label{tab1}
\end{center}
{\footnotesize
Left side:\ sample size is $n=200$; Center:\ sample size is $n=500$; Right side:\ sample size is $n=1000$;\\
Logit:\ ordinary logistic regression; CBPS:\ Covariate balancing propensity score\cite{Im2014}; RCAL:\ Calibrating method proposed by Tan\cite{Ta2020}; BLPS:\ Bayesian method proposed by Saarela et al.\cite{Sa2015}; BRCAL:\ Proposed method;\\
{\bf (a)}:\ ``Small complexity and good overlap,'' {\bf (b)}:\ ``Small complexity and poor overlap."
}
\end{table}

\begin{table}[htbp]
\begin{center}
\caption{Summary of estimated causal effects under the scenario {\bf (c)} and {\ (d)} (``Confounder" situation):\ Bias, root mean squared error (RMSE), 95\% coverage probability (95\%CP), average length of CI (AvL), and bias and RMSE ratio (BR/RR) compared to the bias/RMSE of our proposed method of the IPW estimator in 10,000 iterations were summarized for each scenario (in ``Scenario" column) and method (in the ``Method'' column). RCAL:\ CV was used for the CV described in Section 4.2. BRCAL:\ PCIC was used for the PCIC to select the learning late $\omega$ from candidates: $(0.2, 0.5, 1.0, 1.5)$ (see Section 4.2).}
\scalebox{0.63}{
\begin{tabular}{ccc|cccccc|cccccc|cccccc}\hline
{\bf Scenario}&\multicolumn{2}{c|}{\bf Method}&\begin{tabular}{c}{\bf Bias}\\$(\times 10^{-2})$\end{tabular}&\begin{tabular}{c}{\bf RMSE}\\$(\times 10^{-2})$\end{tabular}&{\bf 95\%CP}&{\bf AvL}&{\bf BR}&{\bf RR}&\begin{tabular}{c}{\bf Bias}\\$(\times 10^{-2})$\end{tabular}&\begin{tabular}{c}{\bf RMSE}\\$(\times 10^{-2})$\end{tabular}&{\bf 95\%CP}&{\bf AvL}&{\bf BR}&{\bf RR}&\begin{tabular}{c}{\bf Bias}\\$(\times 10^{-2})$\end{tabular}&\begin{tabular}{c}{\bf RMSE}\\$(\times 10^{-2})$\end{tabular}&{\bf 95\%CP}&{\bf AvL}&{\bf BR}&{\bf RR}\\
\hline
(c) & \multicolumn{2}{c|}{Logit} & 0.790 & 5.498 & 0.945 & 0.211 & 1.211 & 1.000 & 0.743 & 3.425 & 0.951 & 0.134 & 1.071 & 0.998 & 0.505 & 2.447 & 0.945 & 0.095 & 1.071 & 0.998 \\ \cline{2-21}
 & \multicolumn{2}{c|}{CBPS} & 0.852 & 5.519 & 0.944 & 0.212 & 1.306 & 1.004 & 0.762 & 3.436 & 0.950 & 0.135 & 1.099 & 1.002 & 0.501 & 2.454 & 0.945 & 0.095 & 1.064 & 1.001 \\ \cline{2-21}
 & RCAL & CV & 0.695 & 5.347 & 0.950 & 0.209 & 1.066 & 0.973 & 0.209 & 3.377 & 0.950 & 0.132 & 0.301 & 0.984 & 0.227 & 2.421 & 0.943 & 0.093 & 0.482 & 0.988 \\ \cline{3-21}
 &  & $\lambda=0.01$ & 0.029 & 5.387 & 0.949 & 0.209 & -- & -- & 0.006 & 3.352 & 0.952 & 0.132 & -- & -- & 0.134 & 2.398 & 0.946 & 0.093 & -- & -- \\ \cline{3-21}
 &  & $\lambda=0.05$ & 2.499 & 5.220 & 0.955 & 0.208 & -- & -- & 2.798 & 3.276 & 0.954 & 0.131 & -- & -- & 2.915 & 2.364 & 0.948 & 0.093 & -- & -- \\ \cline{3-21}
 &  & $\lambda=0.10$ & 4.246 & 5.217 & 0.953 & 0.207 & -- & -- & 4.471 & 3.315 & 0.951 & 0.131 & -- & -- & 4.571 & 2.421 & 0.942 & 0.092 & -- & -- \\ \cline{2-21}
 & \multicolumn{2}{c|}{BLPS} & 1.030 & 5.533 & 0.990 & 0.278 & -- & -- & 0.835 & 3.432 & 0.990 & 0.176 & -- & -- & 0.547 & 2.450 & 0.989 & 0.124 & -- & -- \\ \cline{2-21}
 & BRCAL & PCIC & 0.652 & 5.495 & 0.965 & 0.251 & Ref. & Ref. & 0.694 & 3.430 & 0.967 & 0.146 & Ref. & Ref. & 0.471 & 2.451 & 0.963 & 0.105 & Ref. & Ref. \\ \cline{3-21}
 &  & $\omega=0.2$ & 0.339 & 5.378 & 1.000 & 0.619 & -- & -- & 0.231 & 3.393 & 1.000 & 0.392 & -- & -- & 0.269 & 2.438 & 1.000 & 0.277 & -- & -- \\ \cline{3-21}
 &  & $\omega=0.5$ & 0.367 & 5.455 & 1.000 & 0.392 & -- & -- & 0.568 & 3.420 & 1.000 & 0.248 & -- & -- & 0.413 & 2.448 & 1.000 & 0.175 & -- & -- \\ \cline{3-21}
 &  & $\omega=1.0$ & 0.605 & 5.485 & 0.990 & 0.277 & -- & -- & 0.661 & 3.428 & 0.989 & 0.175 & -- & -- & 0.457 & 2.450 & 0.990 & 0.124 & -- & -- \\ \cline{3-21}
 &  & $\omega=1.5$ & 0.692 & 5.497 & 0.962 & 0.226 & -- & -- & 0.699 & 3.431 & 0.964 & 0.143 & -- & -- & 0.474 & 2.451 & 0.959 & 0.101 & -- & -- \\ \hline
(d) & \multicolumn{2}{c|}{Logit} & 0.568 & 5.562 & 0.945 & 0.215 & 1.565 & 1.000 & 1.033 & 3.547 & 0.944 & 0.137 & 1.143 & 0.998 & 1.013 & 2.467 & 0.950 & 0.097 & 1.124 & 0.997 \\ \cline{2-21}
 & \multicolumn{2}{c|}{CBPS} & 0.531 & 5.587 & 0.947 & 0.216 & 1.463 & 1.005 & 0.954 & 3.561 & 0.944 & 0.138 & 1.056 & 1.002 & 0.926 & 2.476 & 0.951 & 0.097 & 1.027 & 1.001 \\ \cline{2-21}
 & RCAL & CV & 0.785 & 5.361 & 0.948 & 0.208 & 2.165 & 0.964 & 0.692 & 3.476 & 0.941 & 0.132 & 0.766 & 0.978 & 0.918 & 2.429 & 0.947 & 0.093 & 1.019 & 0.982 \\ \cline{3-21}
 &  & $\lambda=0.01$ & 0.068 & 5.418 & 0.946 & 0.209 & -- & -- & 0.536 & 3.447 & 0.942 & 0.132 & -- & -- & 0.593 & 2.396 & 0.948 & 0.093 & -- & -- \\ \cline{3-21}
 &  & $\lambda=0.05$ & 2.540 & 5.200 & 0.954 & 0.208 & -- & -- & 2.084 & 3.329 & 0.948 & 0.131 & -- & -- & 1.992 & 2.322 & 0.955 & 0.092 & -- & -- \\ \cline{3-21}
 &  & $\lambda=0.10$ & 5.092 & 5.177 & 0.954 & 0.206 & -- & -- & 5.053 & 3.381 & 0.943 & 0.130 & -- & -- & 5.200 & 2.423 & 0.943 & 0.092 & -- & -- \\ \cline{2-21}
 & \multicolumn{2}{c|}{BLPS} & 0.869 & 5.611 & 0.988 & 0.278 & -- & -- & 1.156 & 3.558 & 0.985 & 0.176 & -- & -- & 1.077 & 2.470 & 0.988 & 0.124 & -- & -- \\ \cline{2-21}
 & BRCAL & PCIC & 0.363 & 5.560 & 0.958 & 0.236 & Ref. & Ref. & 0.904 & 3.554 & 0.957 & 0.145 & Ref. & Ref. & 0.901 & 2.473 & 0.962 & 0.104 & Ref. & Ref. \\ \cline{3-21}
 &  & $\omega=0.2$ & 0.630 & 5.420 & 1.000 & 0.619 & -- & -- & 0.516 & 3.512 & 1.000 & 0.392 & -- & -- & 0.748 & 2.458 & 1.000 & 0.277 & -- & -- \\ \cline{3-21}
 &  & $\omega=0.5$ & 0.098 & 5.513 & 0.999 & 0.392 & -- & -- & 0.791 & 3.543 & 1.000 & 0.248 & -- & -- & 0.871 & 2.469 & 0.999 & 0.175 & -- & -- \\ \cline{3-21}
 &  & $\omega=1.0$ & 0.316 & 5.549 & 0.988 & 0.277 & -- & -- & 0.877 & 3.549 & 0.985 & 0.175 & -- & -- & 0.893 & 2.472 & 0.987 & 0.124 & -- & -- \\ \cline{3-21}
 &  & $\omega=1.5$ & 0.389 & 5.562 & 0.957 & 0.227 & -- & -- & 0.905 & 3.554 & 0.954 & 0.143 & -- & -- & 0.900 & 2.473 & 0.958 & 0.101 & -- & -- \\ \hline
\end{tabular}
}
\label{tab2}
\end{center}
{\footnotesize
Left side:\ sample size is $n=200$; Center:\ sample size is $n=500$; Right side:\ sample size is $n=1000$;\\
Logit:\ ordinary logistic regression; CBPS:\ Covariate balancing propensity score\cite{Im2014}; RCAL:\ Calibrating method proposed by Tan\cite{Ta2020}; BLPS:\ Bayesian method proposed by Saarela et al.\cite{Sa2015}; BRCAL:\ Proposed method;\\
{\bf (c)}:\ ``Moderate complexity and good overlap,'' {\bf (d)}:\ ``Moderate complexity and poor overlap.''
}
\end{table}

\begin{table}[htbp]
\begin{center}
\caption{Summary of estimated causal effects under the scenario {\bf (e)} and {\bf (f)} (``Confounder" situation):\ Bias, root mean squared error (RMSE), 95\% coverage probability (95\%CP), average length of CI (AvL), and bias and RMSE ratio (BR/RR) compared to the bias/RMSE of our proposed method of the IPW estimator in 10,000 iterations were summarized for each scenario (in ``Scenario" column) and method (in the ``Method'' column). RCAL:\ CV was used for the CV described in Section 4.2. BRCAL:\ PCIC was used for the PCIC to select the learning late $\omega$ from candidates: $(0.2, 0.5, 1.0, 1.5)$ (see Section 4.2).}
\scalebox{0.63}{
\begin{tabular}{ccc|cccccc|cccccc|cccccc}\hline
{\bf Scenario}&\multicolumn{2}{c|}{\bf Method}&\begin{tabular}{c}{\bf Bias}\\$(\times 10^{-2})$\end{tabular}&\begin{tabular}{c}{\bf RMSE}\\$(\times 10^{-2})$\end{tabular}&{\bf 95\%CP}&{\bf AvL}&{\bf BR}&{\bf RR}&\begin{tabular}{c}{\bf Bias}\\$(\times 10^{-2})$\end{tabular}&\begin{tabular}{c}{\bf RMSE}\\$(\times 10^{-2})$\end{tabular}&{\bf 95\%CP}&{\bf AvL}&{\bf BR}&{\bf RR}&\begin{tabular}{c}{\bf Bias}\\$(\times 10^{-2})$\end{tabular}&\begin{tabular}{c}{\bf RMSE}\\$(\times 10^{-2})$\end{tabular}&{\bf 95\%CP}&{\bf AvL}&{\bf BR}&{\bf RR}\\
\hline
(e) & \multicolumn{2}{c|}{Logit} & 2.293 & 6.386 & 0.935 & 0.238 & 1.106 & 0.999 & 1.584 & 3.942 & 0.945 & 0.153 & 1.238 & 0.995 & 1.288 & 2.795 & 0.945 & 0.108 & 1.318 & 0.994 \\ \cline{2-21}
 & \multicolumn{2}{c|}{CBPS} & 2.015 & 6.430 & 0.931 & 0.239 & 0.972 & 1.006 & 1.250 & 3.970 & 0.946 & 0.154 & 0.977 & 1.003 & 0.957 & 2.817 & 0.946 & 0.109 & 0.979 & 1.001 \\ \cline{2-21}
 & RCAL & CV & 2.584 & 6.173 & 0.907 & 0.208 & 1.246 & 0.965 & 1.467 & 3.994 & 0.903 & 0.132 & 1.148 & 1.008 & 0.982 & 2.881 & 0.891 & 0.093 & 1.004 & 1.024 \\ \cline{3-21}
 &  & $\lambda=0.01$ & 2.880 & 6.704 & 0.886 & 0.209 & -- & -- & 1.455 & 3.986 & 0.905 & 0.132 & -- & -- & 0.960 & 2.815 & 0.900 & 0.093 & -- & -- \\ \cline{3-21}
 &  & $\lambda=0.05$ & 2.754 & 5.793 & 0.927 & 0.208 & -- & -- & 2.278 & 3.586 & 0.934 & 0.131 & -- & -- & 1.964 & 2.548 & 0.931 & 0.093 & -- & -- \\ \cline{3-21}
 &  & $\lambda=0.10$ & 3.970 & 5.414 & 0.941 & 0.205 & -- & -- & 4.223 & 3.407 & 0.942 & 0.129 & -- & -- & 4.204 & 2.462 & 0.933 & 0.091 & -- & -- \\ \cline{2-21}
 & \multicolumn{2}{c|}{BLPS} & 2.235 & 6.525 & 0.970 & 0.281 & -- & -- & 1.545 & 3.972 & 0.974 & 0.177 & -- & -- & 1.265 & 2.806 & 0.974 & 0.125 & -- & -- \\ \cline{2-21}
 & BRCAL & PCIC & 2.073 & 6.393 & 0.927 & 0.229 & Ref. & Ref. & 1.279 & 3.960 & 0.935 & 0.146 & Ref. & Ref. & 0.977 & 2.813 & 0.932 & 0.105 & Ref. & Ref. \\ \cline{3-21}
 &  & $\omega=0.2$ & 2.388 & 6.159 & 1.000 & 0.621 & -- & -- & 1.475 & 3.904 & 1.000 & 0.392 & -- & -- & 1.068 & 2.793 & 1.000 & 0.277 & -- & -- \\ \cline{3-21}
 &  & $\omega=0.5$ & 2.180 & 6.318 & 0.998 & 0.393 & -- & -- & 1.325 & 3.945 & 0.998 & 0.248 & -- & -- & 0.992 & 2.808 & 0.998 & 0.175 & -- & -- \\ \cline{3-21}
 &  & $\omega=1.0$ & 2.103 & 6.373 & 0.970 & 0.278 & -- & -- & 1.291 & 3.956 & 0.973 & 0.175 & -- & -- & 0.975 & 2.813 & 0.973 & 0.124 & -- & -- \\ \cline{3-21}
 &  & $\omega=1.5$ & 2.071 & 6.395 & 0.926 & 0.227 & -- & -- & 1.278 & 3.961 & 0.930 & 0.143 & -- & -- & 0.976 & 2.813 & 0.924 & 0.101 & -- & -- \\ \hline
(f) & \multicolumn{2}{c|}{Logit} & 3.451 & 6.691 & 0.928 & 0.246 & 1.283 & 0.971 & 3.568 & 4.103 & 0.942 & 0.158 & 1.321 & 0.971 & 3.438 & 2.958 & 0.940 & 0.112 & 1.339 & 0.973 \\ \cline{2-21}
 & \multicolumn{2}{c|}{CBPS} & 2.625 & 6.933 & 0.925 & 0.254 & 0.975 & 1.006 & 2.672 & 4.238 & 0.943 & 0.165 & 0.989 & 1.003 & 2.553 & 3.044 & 0.942 & 0.117 & 0.995 & 1.001 \\ \cline{2-21}
 & RCAL & CV & 4.529 & 6.695 & 0.881 & 0.208 & 1.683 & 0.971 & 3.798 & 4.446 & 0.867 & 0.132 & 1.406 & 1.052 & 3.477 & 3.318 & 0.844 & 0.093 & 1.355 & 1.091 \\ \cline{3-21}
 &  & $\lambda=0.01$ & 5.853 & 8.502 & 0.818 & 140.571 & -- & -- & 3.918 & 4.642 & 0.849 & 0.132 & -- & -- & 3.387 & 3.284 & 0.848 & 0.093 & -- & -- \\ \cline{3-21}
 &  & $\lambda=0.05$ & 4.356 & 6.537 & 0.888 & 0.268 & -- & -- & 4.011 & 3.907 & 0.906 & 0.132 & -- & -- & 3.814 & 2.837 & 0.896 & 0.093 & -- & -- \\ \cline{3-21}
 &  & $\lambda=0.10$ & 5.370 & 5.729 & 0.926 & 0.206 & -- & -- & 5.703 & 3.570 & 0.929 & 0.130 & -- & -- & 5.570 & 2.649 & 0.916 & 0.092 & -- & -- \\ \cline{2-21}
 & \multicolumn{2}{c|}{BLPS} & 3.274 & 6.868 & 0.964 & 0.284 & -- & -- & 3.500 & 4.142 & 0.970 & 0.178 & -- & -- & 3.402 & 2.971 & 0.964 & 0.126 & -- & -- \\ \cline{2-21}
 & BRCAL & PCIC & 2.691 & 6.891 & 0.904 & 0.231 & Ref. & Ref. & 2.701 & 4.227 & 0.918 & 0.148 & Ref. & Ref. & 2.567 & 3.040 & 0.918 & 0.107 & Ref. & Ref. \\ \cline{3-21}
 &  & $\omega=0.2$ & 3.165 & 6.639 & 1.000 & 0.620 & -- & -- & 2.905 & 4.160 & 1.000 & 0.391 & -- & -- & 2.648 & 3.019 & 1.000 & 0.276 & -- & -- \\ \cline{3-21}
 &  & $\omega=0.5$ & 2.815 & 6.812 & 0.995 & 0.392 & -- & -- & 2.765 & 4.207 & 0.997 & 0.247 & -- & -- & 2.602 & 3.035 & 0.997 & 0.175 & -- & -- \\ \cline{3-21}
 &  & $\omega=1.0$ & 2.727 & 6.872 & 0.955 & 0.277 & -- & -- & 2.713 & 4.222 & 0.962 & 0.175 & -- & -- & 2.570 & 3.039 & 0.956 & 0.124 & -- & -- \\ \cline{3-21}
 &  & $\omega=1.5$ & 2.691 & 6.893 & 0.900 & 0.226 & -- & -- & 2.695 & 4.227 & 0.909 & 0.143 & -- & -- & 2.569 & 3.041 & 0.905 & 0.101 & -- & -- \\ \hline
\end{tabular}
}
\label{tab3}
\end{center}
{\footnotesize
Left side:\ sample size is $n=200$; Center:\ sample size is $n=500$; Right side:\ sample size is $n=1000$;\\
Logit:\ ordinary logistic regression; CBPS:\ Covariate balancing propensity score\cite{Im2014}; RCAL:\ Calibrating method proposed by Tan\cite{Ta2020}; BLPS:\ Bayesian method proposed by Saarela et al.\cite{Sa2015}; BRCAL:\ Proposed method;\\
{\bf (e)}:\ ``Large complexity and good overlap,'' {\bf (f)}:\ ``Large complexity and poor overlap.''
}
\end{table}
\end{landscape}

\begin{figure}[H]
\begin{center}
\begin{tabular}{c}
\includegraphics[width=18cm]{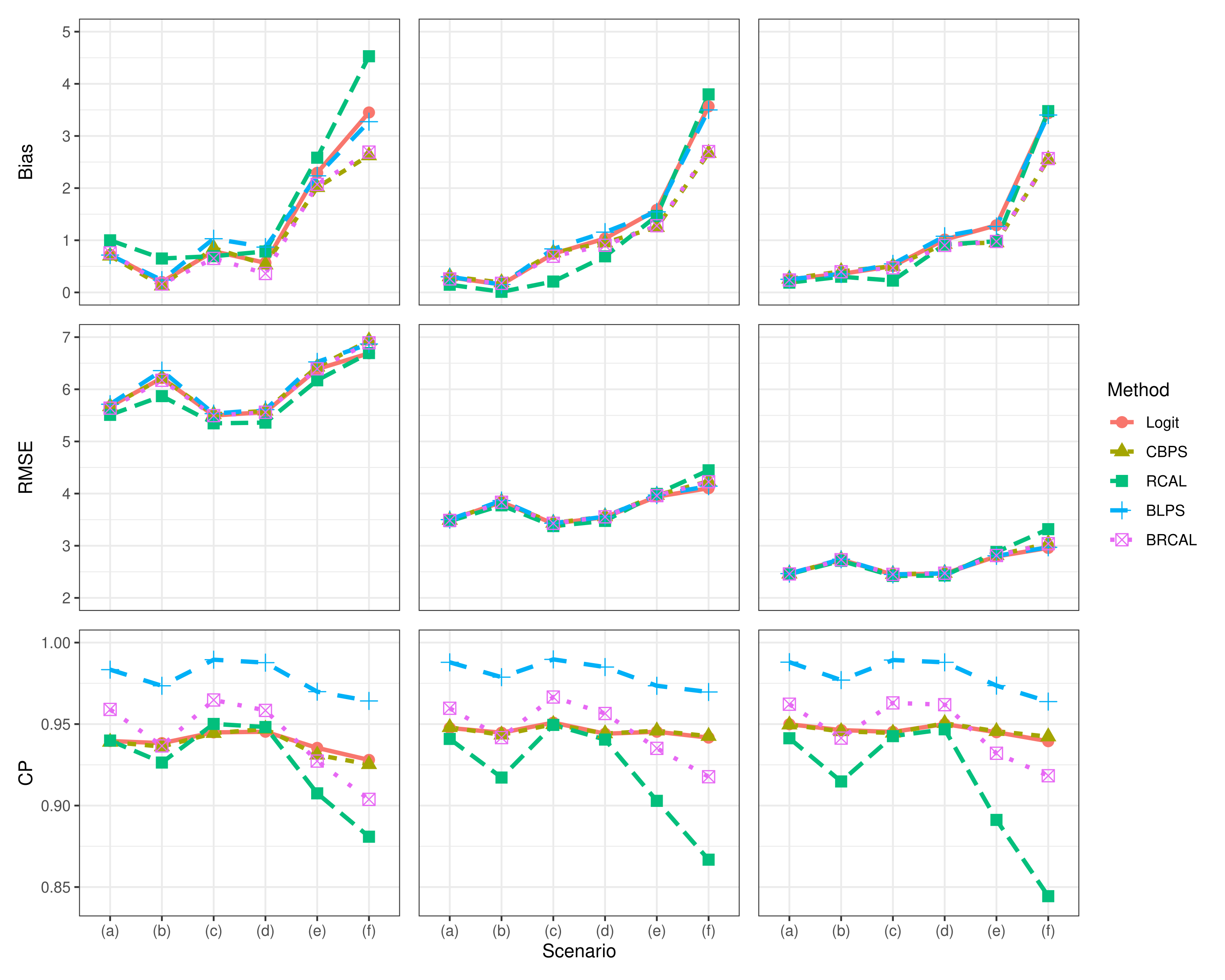}
\end{tabular}
\caption{Summary of estimated causal effects:\ Bias, root mean squared error (RMSE), and 95\% coverage probability (95\%CP) in 10,000 iterations were summarized under ``Covariate" situation. The left panel: $n=200$; the center panel: $n=500$; The right panel: $n=1000$. RCAL was used for the CV described in Section 4.2. BRCAL was used for the PCIC to select the learning rate $\omega$ from candidates: $(0.2, 0.5, 1.0, 1.5)$ (see Section 4.2).}
\label{fig1}
\end{center}
{\footnotesize
{\bf (a)}:\ ``Small complexity and good overlap,'' {\bf (b)}:\ ``Small complexity and poor overlap,'' {\bf (c)}:\ ``Moderate complexity and good overlap,'' {\bf (d)}:\ ``Moderate complexity and poor overlap,'' {\bf (e)}:\ ``Large complexity and good overlap,'' {\bf (f)}:\ ``Large complexity and poor overlap.''
}
\end{figure}

\newpage
\appendix

\renewcommand{\thetable}{\Alph{section}.\arabic{table}}
\renewcommand{\thefigure}{\Alph{section}.\arabic{figure}}
\setcounter{table}{0}
\setcounter{figure}{0}

\section{Mathematical Calculations}
\label{appA}
\subsection{Derivation of Loss Functions}
First, we consider the loss function for the propensity score (\ref{eq1}). Specifically, we consider the first derivatives respect to $\bld{\alpha}$:
\begin{align*}
\frac{\partial}{\partial\bld{\alpha}}\ell_{A}(\bld{\alpha})&=\sum_{i=1}^{n}\left[-A_{i}\exp\left\{-\bld{\alpha}^{\top}g(\bld{X}_{i})\right\}g(\bld{X}_{i})+(1-A_{i})g(\bld{X}_{i})\right.\\
&\hspace{0.5cm}\left.+(1-A_{i})\exp\left\{\bld{\alpha}^{\top}g(\bld{X}_{i})\right\}g(\bld{X}_{i})-A_{i}g(\bld{X}_{i})\right]\\
&=\sum_{i=1}^{n}\left[-A_{i}\left\{1+\exp\left\{-\bld{\alpha}^{\top}g(\bld{X}_{i})\right\}\right\}+(1-A_{i})\left\{1+\exp\left\{\bld{\alpha}^{\top}g(\bld{X}_{i})\right\}\right\}\right]g(\bld{X}_{i})\\
&=\sum_{i=1}^{n}\left[-\frac{A_{i}}{e(\bld{X}_{i};\bld{\alpha})}+\frac{1-A_{i}}{1-e(\bld{X}_{i};\bld{\alpha})}\right]g(\bld{X}_{i}),
\end{align*}
where we use the relationship from the propensity score model (i.e., logistic regression model) for the final equality. Additionally, for the penalty term (\ref{pen1}),
$$
\frac{\partial}{\partial\bld{\alpha}}\lambda\|\bld{\alpha}\|_{1}=\lambda\times sgn(\bld{\alpha})=\left\{
\begin{array}{lc}
\lambda&\alpha_{\ell}>0\\
-\lambda&\alpha_{\ell}<0\\
\left[-\lambda,\lambda\right]&\alpha_{\ell}=0\\
\end{array}
\right.
$$
From the results, the inequality (\ref{ineq1}) can be obtained.

Next, we consider the loss functions for the causal effect; here, we only consider the loss function (\ref{eq3}) (i.e., for the treatment group). Considering the first derivatives respect to $\theta_{1}$,
$$
\frac{\partial}{\partial\theta_{1}}\ell_{Y_{1}}(\theta_{1},\bld{\alpha})=-2\sum_{i=1}^{n}\frac{A_{i}}{e_{i}(\bld{\alpha})}\left(Y_{i}-\theta_{1}\right).
$$
Therefore, a part of IPW estimator (\ref{IPW_est}) can be obtained.

\subsection{Derivation of Posterior Distributions}
The posterior distributions for $\theta_{k}$ are derived as follows:
\begin{align*}
p(\theta_k \mid \bm{Y}, \bm{X}, \bm{A}, \bm{\alpha}) &\propto \exp\left\{ -\frac{\tau_k}{2} (\theta_k-\mu_k)^2 \right\} \exp\left\{ -\omega \sum_{i=1}^n \frac{A_i^k(1-A_i)^{1-k}}{e_i(\bm{\alpha})^k \{1-e_i(\bm{\alpha})\}^{1-k} } (Y_{i}-\theta_k)^2 \right\} \\
&= \exp\left\{ -\frac{\tau_k}{2} (\theta_k-\mu_k)^2 \right\} \exp\left\{ -\frac{\omega}{2} \sum_{i=1}^n s_{ki} (Y_{i}-\theta_k)^2 \right\} \\
&= \exp\left\{ -\frac{1}{2}\left[\tau_k (\theta_k^2-2\mu_k\theta_k+\mu_k^2) -\omega \sum_{i=1}^n s_{ki} (Y_{i}^2-2\theta_kY_{i}+\theta_k^2)\right] \right\} \\
&\propto \exp\left\{ -\frac{1}{2}\left[\left(\tau_k+\omega\sum_{i=1}^n s_{ki}\right)\theta_k^2-2\left(\tau_k\mu_k+\omega\sum_{i=1}^n s_{ki}Y_{i}\right)\theta_k\right] \right\} \\
&\propto \exp\left\{ -\frac{\tau_k+\omega\sum_{i=1}^n s_{ki}}{2}\left(\theta_k-\frac{\tau_k\mu_k+\omega\sum_{i=1}^n s_{ki}Y_{i}}{\tau_k+\omega\sum_{i=1}^n s_{ki}}\right)^2 \right\} \\
&= \exp\left\{ -\frac{\Tilde{\tau}_k}{2} (\theta_k - \Tilde{\mu}_k)^2 \ \right\},
\end{align*}
where ``$\propto$" denotes proportionality with respect to $\theta_{k}$, and
\begin{equation*}
s_{ki} = \frac{2 A_i^k(1-A_i)^{1-k}}{e_i(\bm{\alpha})^k \{1-e_i(\bm{\alpha})\}^{1-k} }, ~~ \Tilde{\mu}_k = \Tilde{\tau}_k^{-1} \left(\tau_k\mu_k + \omega\sum_{i=1}^n s_{ki}Y_i \right) ~~ \mbox{and} ~~ \Tilde{\tau}_k = \tau_k + \omega\sum_{i=1}^n s_{ki}.
\end{equation*}

\newpage
\section{Additional Materials for Simulation Experiments}
\label{appB}
\subsection{Details for Data-generating Mechanism}
For each situation, the linear predictor $h_{A}\left(\bld{X}_{i}\right)$ of the true propensity score model was set as follows:
\begin{description}
\item[(a)] Small complexity and good overlap
$$
h_{A}\left(\bld{X}_{i}\right)=\left(X_{i1},\dots,X_{i7}\right)\bld{\beta}_{11},\ \bld{\beta}_{11}=\left(0.4,0.8,-0.25,0.6,-0.4,-0.8,-0.5,0.7\right)^{\top}.
$$
\item[(b)] Small complexity and poor overlap
$$
h_{A}\left(\bld{X}_{i}\right)=\left(X_{i1},\dots,X_{i7}\right)\bld{\beta}_{12},\ \bld{\beta}_{12}=2.5\bld{\beta}_{11},
$$
\item[(c)] Moderate complexity and good overlap
$$
h_{A}\left(\bld{X}_{i}\right)=\left(X_{i1},\dots,X_{i7},X_{i2}^2,X_{i1}\times X_{i3},X_{i2}\times X_{i4},X_{i4}\times X_{i5},X_{i5}\times X_{i6}\right)\bld{\beta}_{21},
$$
$$
\bld{\beta}_{21}=\left(0.6\bld{\beta}_{11}^{\top},1,0.96,-0.3,-0.48,-0.96\right)^{\top}.
$$
\item[(d)] Moderate complexity and poor overlap
$$
h_{A}\left(\bld{X}_{i}\right)=\left(X_{i1},\dots,X_{i7},X_{i2}^2,X_{i1}\times X_{i3},X_{i2}\times X_{i4},X_{i4}\times X_{i5},X_{i5}\times X_{i6}\right)\bld{\beta}_{22},
$$
$$
\bld{\beta}_{22}=\left(0.4\bld{\beta}_{11}^{\top},1,1.6,-0.5,-0.8,-1.6\right)^{\top}.
$$
\item[(e)] Large complexity and good overlap
\begin{align*}
h_{A}\left(\bld{X}_{i}\right)&=\left(X_{i1},\dots,X_{i7},X_{i1}\times X_{i3},X_{i5}\times X_{i6},sin\left(2(X_{i2}\times X_{i4})\right),cos\left(2(X_{i4}\times X_{i5})\right),\right.\\
&\hspace{0.5cm}\left.\exp\left(2(X_{i2}\times X_{i4})\right),X_{i2}\times X_{i5}\times X_{i6}\right)\bld{\beta}_{31},
\end{align*}
$$
\bld{\beta}_{31}=\left(\bld{\beta}_{11}^{\top},0.4,-0.4,0.5,0.5,-0.25,-0.5\right)^{\top}.
$$
\item[(f)] Large complexity and poor overlap
\begin{align*}
h_{A}\left(\bld{X}_{i}\right)&=\left(X_{i1},\dots,X_{i7},X_{i1}\times X_{i3},X_{i5}\times X_{i6},sin\left(2(X_{i2}\times X_{i4})\right),cos\left(2(X_{i4}\times X_{i5})\right),\right.\\
&\hspace{0.5cm}\left.\exp\left(2(X_{i2}\times X_{i4})\right),X_{i2}\times X_{i5}\times X_{i6}\right)\bld{\beta}_{32},
\end{align*}
$$
\bld{\beta}_{32}=\left(0.5\bld{\beta}_{11}^{\top},0.8,-0.8,1,1,-0.5,-1\right)^{\top}.
$$
\end{description}

\subsection{Additional Results for Simulation Experiments}
In this section, we additionally summarize the results under ``Predictors" situations. Except for the bias results, a similar tendency is observed under ``Confounders" situations. Furthermore, in scenario {\bf (b)}, there are unexpected results in terms of RMSE and 95\% CP.

The BRCAL and CBPS exhibit almost the same value of bias and RMSE in almost all scenarios. In contrast, the RCAL exhibits large, unstable bias compared to other methods. The results of all methods become consistent in large samples.

\begin{landscape}
\begin{table}[htbp]
\begin{center}
\caption{Summary of estimated causal effects under the scenario {\bf (a)} and {\bf (b)} (``Predictor" situation):\ Bias, root mean squared error (RMSE), 95\% coverage probability (95\%CP), average length of CI (AvL), and bias and RMSE ratio (BR/RR) compared to the bias/RMSE of our proposed method of the IPW estimator in 10,000 iterations were summarized for each scenario (in ``Scenario" column) and method (in the ``Method'' column). RCAL:\ CV was used for the CV described in Section 4.2. BRCAL:\ PCIC was used for the PCIC to select the learning rate $\omega$ from candidates: $(0.2, 0.5, 1.0, 1.5)$ (see Section 4.2).}
\scalebox{0.63}{
\begin{tabular}{ccc|cccccc|cccccc|cccccc}\hline
{\bf Scenario}&\multicolumn{2}{c|}{\bf Method}&\begin{tabular}{c}{\bf Bias}\\$(\times 10^{-2})$\end{tabular}&\begin{tabular}{c}{\bf RMSE}\\$(\times 10^{-2})$\end{tabular}&{\bf 95\%CP}&{\bf AvL}&{\bf BR}&{\bf RR}&\begin{tabular}{c}{\bf Bias}\\$(\times 10^{-2})$\end{tabular}&\begin{tabular}{c}{\bf RMSE}\\$(\times 10^{-2})$\end{tabular}&{\bf 95\%CP}&{\bf AvL}&{\bf BR}&{\bf RR}&\begin{tabular}{c}{\bf Bias}\\$(\times 10^{-2})$\end{tabular}&\begin{tabular}{c}{\bf RMSE}\\$(\times 10^{-2})$\end{tabular}&{\bf 95\%CP}&{\bf AvL}&{\bf BR}&{\bf RR}\\
\hline
(a) & \multicolumn{2}{c|}{Logit} & 0.183 & 6.426 & 0.930 & 0.232 & 0.835 & 1.023 & 0.134 & 3.928 & 0.945 & 0.150 & 0.956 & 1.010 & 0.105 & 2.764 & 0.946 & 0.107 & 0.789 & 1.005 \\ \cline{2-21} 
 & \multicolumn{2}{c|}{CBPS} & 0.283 & 6.345 & 0.928 & 0.230 & 1.295 & 1.010 & 0.184 & 3.903 & 0.943 & 0.149 & 1.314 & 1.004 & 0.152 & 2.757 & 0.944 & 0.106 & 1.145 & 1.002 \\ \cline{2-21}
 & RCAL & CV & 0.556 & 5.751 & 0.932 & 0.208 & 2.541 & 0.915 & 0.337 & 3.779 & 0.919 & 0.132 & 2.412 & 0.972 & 0.134 & 2.704 & 0.917 & 0.093 & 1.007 & 0.983 \\ \cline{3-21}
 &  & $\lambda=0.01$ & 0.548 & 6.448 & 0.903 & 0.208 & -- & -- & 0.300 & 3.795 & 0.919 & 0.132 & -- & -- & 0.190 & 2.642 & 0.923 & 0.093 & -- & -- \\ \cline{3-21}
 &  & $\lambda=0.05$ & 0.514 & 5.440 & 0.944 & 0.209 & -- & -- & 0.866 & 3.392 & 0.946 & 0.131 & -- & -- & 1.011 & 2.380 & 0.948 & 0.093 & -- & -- \\ \cline{3-21}
 &  & $\lambda=0.10$ & 0.362 & 5.236 & 0.950 & 0.207 & -- & -- & 0.399 & 3.303 & 0.953 & 0.130 & -- & -- & 0.321 & 2.328 & 0.951 & 0.092 & -- & -- \\ \cline{2-21}
 & \multicolumn{2}{c|}{BLPS} & 0.217 & 6.771 & 0.969 & 0.286 & -- & -- & 0.142 & 3.999 & 0.973 & 0.177 & -- & -- & 0.113 & 2.788 & 0.973 & 0.125 & -- & -- \\ \cline{2-21}
 & BRCAL & PCIC & 0.219 & 6.284 & 0.930 & 0.229 & Ref. & Ref. & 0.140 & 3.889 & 0.936 & 0.145 & Ref. & Ref. & 0.133 & 2.751 & 0.936 & 0.103 & Ref. & Ref. \\ \cline{2-21}
 &  & $\omega=0.2$ & 0.110 & 5.959 & 1.000 & 0.621 & -- & -- & 0.107 & 3.805 & 1.000 & 0.392 & -- & -- & 0.005 & 2.720 & 1.000 & 0.277 & -- & -- \\ \cline{3-21}
 &  & $\omega=0.5$ & 0.107 & 6.172 & 0.999 & 0.393 & -- & -- & 0.051 & 3.863 & 0.999 & 0.248 & -- & -- & 0.087 & 2.740 & 0.999 & 0.175 & -- & -- \\ \cline{3-21}
 &  & $\omega=1.0$ & 0.198 & 6.258 & 0.973 & 0.278 & -- & -- & 0.113 & 3.883 & 0.974 & 0.176 & -- & -- & 0.118 & 2.749 & 0.975 & 0.124 & -- & -- \\ \cline{3-21}
 &  & $\omega=1.5$ & 0.219 & 6.287 & 0.928 & 0.227 & -- & -- & 0.140 & 3.889 & 0.935 & 0.143 & -- & -- & 0.133 & 2.751 & 0.934 & 0.101 & -- & -- \\ 
   \hline

(b) & \multicolumn{2}{c|}{Logit} & 0.130 & 11.230 & 0.874 & 0.289 & 0.720 & 1.057 & 0.680 & 7.998 & 0.915 & 0.226 & 1.070 & 1.129 & 0.914 & 6.297 & 0.930 & 0.184 & 1.067 & 1.195 \\ \cline{2-21}
 & \multicolumn{2}{c|}{CBPS} & 0.245 & 10.856 & 0.958 & 0.520 & 1.351 & 1.022 & 0.678 & 7.147 & 0.915 & 0.248 & 1.067 & 1.009 & 0.868 & 5.293 & 0.903 & 0.175 & 1.013 & 1.005 \\ \cline{2-21}
 & RCAL & CV & 1.494 & 6.198 & 0.920 & 0.211 & 8.239 & 0.583 & 0.655 & 5.477 & 0.843 & 0.133 & 1.031 & 0.773 & 0.408 & 4.987 & -- & -- & 0.476 & 0.947 \\ \cline{3-21}
 &  & $\lambda=0.01$ & 6.944 & 20.351 & 0.672 & 10.920 & -- & -- & 2.498 & 11.433 & 0.615 & 0.147 & -- & -- & 1.404 & 7.265 & 0.625 & 0.093 & -- & -- \\ \cline{3-21}
 &  & $\lambda=0.05$ & 1.430 & 8.058 & 0.869 & 0.211 & -- & -- & 0.460 & 4.314 & 0.901 & 0.135 & -- & -- & 0.214 & 2.872 & 0.906 & 0.095 & -- & -- \\ \cline{3-21}
 &  & $\lambda=0.10$ & 1.601 & 5.389 & 0.948 & 0.210 & -- & -- & 1.329 & 3.354 & 0.953 & 0.132 & -- & -- & 1.380 & 2.371 & 0.950 & 0.093 & -- & -- \\ \cline{2-21}
 & \multicolumn{2}{c|}{BLPS} & 0.319 & 12.837 & 0.877 & 0.342 & -- & -- & 0.754 & 8.612 & 0.833 & 0.202 & -- & -- & 0.978 & 6.603 & 0.792 & 0.138 & -- & -- \\ \cline{2-21}
 & BRCAL & PCIC & 0.181 & 10.624 & 0.832 & 0.351 & Ref. & Ref. & 0.635 & 7.085 & 0.777 & 0.196 & Ref. & Ref. & 0.857 & 5.268 & 0.753 & 0.134 & Ref. & Ref. \\ \cline{3-21}
 &  & $\omega=0.2$ & 0.057 & 10.313 & 0.993 & 0.598 & -- & -- & 0.546 & 6.952 & 0.993 & 0.385 & -- & -- & 0.789 & 5.205 & 0.990 & 0.274 & -- & -- \\ \cline{3-21}
 &  & $\omega=0.5$ & 0.134 & 10.644 & 0.924 & 0.376 & -- & -- & 0.614 & 7.070 & 0.915 & 0.243 & -- & -- & 0.840 & 5.256 & 0.901 & 0.173 & -- & -- \\ \cline{3-21}
 &  & $\omega=1.0$ & 0.196 & 10.751 & 0.794 & 0.265 & -- & -- & 0.652 & 7.110 & 0.782 & 0.172 & -- & -- & 0.859 & 5.275 & 0.761 & 0.122 & -- & -- \\ \cline{3-21}
 &  & $\omega=1.5$ & 0.205 & 10.786 & 0.705 & 0.217 & -- & -- & 0.654 & 7.121 & 0.686 & 0.140 & -- & -- & 0.859 & 5.281 & 0.666 & 0.100 & -- & -- \\ \hline
\end{tabular}
}
\label{tab4}
\end{center}
{\footnotesize
Left side:\ sample size is $n=200$; Center:\ sample size is $n=500$; Right side:\ sample size is $n=1000$;\\
Logit:\ ordinary logistic regression; CBPS:\ Covariate balancing propensity score\cite{Im2014}; RCAL:\ Calibrating method proposed by Tan\cite{Ta2020}; BLPS:\ Bayesian method proposed by Saarela et al.\cite{Sa2015}; BRCAL:\ Proposed method;\\
{\bf (a)}:\ ``Small complexity and good overlap,'' {\bf (b)}:\ ``Small complexity and poor overlap.''
}
\end{table}

\begin{table}[htbp]
\begin{center}
\caption{Summary of estimated causal effects under the scenario {\bf (c)} and {\bf (d)} (``Predictor" situation):\ Bias, root mean squared error (RMSE), 95\% coverage probability (95\%CP), average length of CI (AvL), and bias and RMSE ratio (BR/RR) compared to the bias/RMSE of our proposed method of the IPW estimator in 10,000 iterations were summarized for each scenario (in ``Scenario" column) and method (in the ``Method'' column). RCAL:\ CV was used for the CV described in Section 4.2. BRCAL:\ PCIC was used for the PCIC to select the learning late $\omega$ from candidates: $(0.2, 0.5, 1.0, 1.5)$ (see Section 4.2).}
\scalebox{0.63}{
\begin{tabular}{ccc|cccccc|cccccc|cccccc}\hline
{\bf Scenario}&\multicolumn{2}{c|}{\bf Method}&\begin{tabular}{c}{\bf Bias}\\$(\times 10^{-2})$\end{tabular}&\begin{tabular}{c}{\bf RMSE}\\$(\times 10^{-2})$\end{tabular}&{\bf 95\%CP}&{\bf AvL}&{\bf BR}&{\bf RR}&\begin{tabular}{c}{\bf Bias}\\$(\times 10^{-2})$\end{tabular}&\begin{tabular}{c}{\bf RMSE}\\$(\times 10^{-2})$\end{tabular}&{\bf 95\%CP}&{\bf AvL}&{\bf BR}&{\bf RR}&\begin{tabular}{c}{\bf Bias}\\$(\times 10^{-2})$\end{tabular}&\begin{tabular}{c}{\bf RMSE}\\$(\times 10^{-2})$\end{tabular}&{\bf 95\%CP}&{\bf AvL}&{\bf BR}&{\bf RR}\\
\hline
(c) & \multicolumn{2}{c|}{Logit} & 1.897 & 6.275 & 0.937 & 0.226 & 1.181 & 1.025 & 1.319 & 3.800 & 0.947 & 0.145 & 1.192 & 1.009 & 1.240 & 2.693 & 0.947 & 0.103 & 1.140 & 1.005  \\ \cline{2-21} 
 & \multicolumn{2}{c|}{CBPS} & 1.522 & 6.178 & 0.934 & 0.225 & 0.948 & 1.009 & 1.074 & 3.780 & 0.945 & 0.144 & 0.970 & 1.004 & 1.066 & 2.683 & 0.945 & 0.103 & 0.981 & 1.002 \\ \cline{2-21}
 & RCAL & CV & 2.893 & 5.809 & 0.933 & 0.210 & 1.802 & 0.949 & 1.974 & 3.718 & 0.926 & 0.132 & 1.783 & 0.987 & 1.775 & 2.688 & 0.919 & 0.093 & 1.633 & 1.004 \\ \cline{3-21}
 &  & $\lambda=0.01$ & 2.709 & 6.238 & 0.918 & 0.209 & -- & -- & 1.921 & 3.659 & 0.931 & 0.132 & -- & -- & 1.696 & 2.592 & 0.929 & 0.094 & -- & -- \\ \cline{3-21}
 &  & $\lambda=0.05$ & 3.350 & 5.361 & 0.947 & 0.209 & -- & -- & 3.239 & 3.322 & 0.954 & 0.132 & -- & -- & 3.242 & 2.380 & 0.951 & 0.093 & -- & -- \\ \cline{3-21}
 &  & $\lambda=0.10$ & 4.518 & 5.263 & 0.950 & 0.207 & -- & -- & 4.492 & 3.313 & 0.953 & 0.131 & -- & -- & 4.516 & 2.400 & 0.945 & 0.092 & -- & -- \\ \cline{2-21}
 & \multicolumn{2}{c|}{BLPS} & 1.680 & 6.542 & 0.973 & 0.284 & -- & -- & 1.219 & 3.854 & 0.981 & 0.177 & -- & -- & 1.188 & 2.710 & 0.980 & 0.124 & -- & -- \\ \cline{2-21}
 & BRCAL & PCIC & 1.605 & 6.122 & 0.939 & 0.228 & Ref. & Ref. & 1.107 & 3.766 & 0.944 & 0.144 & Ref. & Ref. & 1.087 & 2.679 & 0.942 & 0.102 & Ref. & Ref. \\ \cline{2-21}
 &  & $\omega=0.2$ & 2.097 & 5.815 & 1.000 & 0.621 & -- & -- & 1.379 & 3.683 & 1.000 & 0.392 & -- & -- & 1.211 & 2.649 & 1.000 & 0.277 & -- & -- \\ \cline{3-21}
 &  & $\omega=0.5$ & 1.757 & 6.016 & 0.999 & 0.393 & -- & -- & 1.185 & 3.737 & 0.999 & 0.248 & -- & -- & 1.118 & 2.670 & 0.999 & 0.175 & -- & -- \\ \cline{3-21}
 &  & $\omega=1.0$ & 1.652 & 6.095 & 0.977 & 0.278 & -- & -- & 1.135 & 3.758 & 0.981 & 0.175 & -- & -- & 1.092 & 2.676 & 0.980 & 0.124 & -- & -- \\ \cline{3-21}
 &  & $\omega=1.5$ & 1.606 & 6.123 & 0.939 & 0.227 & -- & -- & 1.110 & 3.766 & 0.944 & 0.143 & -- & -- & 1.088 & 2.679 & 0.940 & 0.101 & -- & -- \\ 
   \hline

(d) & \multicolumn{2}{c|}{Logit} &  2.422  &  6.435  &  0.932  &  0.232  &  1.151  &  1.020 &  2.901  &  3.980  &  0.943  &  0.150  &  1.085  &  1.009 &  2.493  &  2.778  &  0.945  &  0.107  &  1.076  &  1.004 \\ \cline{2-21}
 & \multicolumn{2}{c|}{CBPS} &  2.047  &  6.365  &  0.932  &  0.232  &  0.973  &  1.009 &  2.653  &  3.961  &  0.941  &  0.150  &  0.993  &  1.004 &  2.299  &  2.773  &  0.944  &  0.107  &  0.992  &  1.002 \\ \cline{2-21}
 & RCAL & CV &  3.249  &  5.985  &  0.927  &  0.210  &  1.544  &  0.949 &  3.719  &  4.031  &  0.902  &  0.132  &  1.392  &  1.022 &  3.421  &  2.882  &  0.895  &  0.093  &  1.477  &  1.041 \\ \cline{3-21}
 &  & $\lambda=0.01$ &  3.945  &  6.914  &  0.888  &  0.210  &  --  &  -- &  3.618  &  3.980  &  0.905  &  0.132  &  --  &  -- &  3.124  &  2.754  &  0.912  &  0.094  &  --  &  -- \\ \cline{3-21}
 &  & $\lambda=0.05$ &  3.039  &  5.423  &  0.948  &  0.210  &  --  &  -- &  3.471  &  3.409  &  0.946  &  0.132  &  --  &  -- &  3.109  &  2.406  &  0.945  &  0.093  &  --  &  -- \\ \cline{3-21}
 &  & $\lambda=0.10$ &  4.553  &  5.227  &  0.952  &  0.207  &  --  &  -- &  5.460  &  3.373  &  0.944  &  0.130  &  --  &  -- &  5.304  &  2.441  &  0.939  &  0.092  &  --  &  -- \\ \cline{2-21}
 & \multicolumn{2}{c|}{BLPS} &  2.210  &  6.757  &  0.969  &  0.285  &  --  &  -- &  2.823  &  4.044  &  0.972  &  0.177  &  --  &  -- &  2.447  &  2.799  &  0.973  &  0.125  &  --  &  -- \\ \cline{2-21}
 & BRCAL & PCIC &  2.104  &  6.308  &  0.932  &  0.227  &  Ref.  &  Ref. &  2.672  &  3.944  &  0.930  &  0.144  &  Ref.  &  Ref. &  2.317  &  2.768  &  0.934  &  0.102  &  Ref.  &  Ref. \\ \cline{3-21}
 &  & $\omega=0.2$ &  2.507  &  5.970  &  1.000  &  0.621  &  --  &  -- &  2.842  &  3.851  &  1.000  &  0.392  &  --  &  -- &  2.404  &  2.736  &  1.000  &  0.277  &  --  &  -- \\ \cline{3-21}
 &  & $\omega=0.5$ &  2.216  &  6.196  &  0.998  &  0.393  &  --  &  -- &  2.715  &  3.915  &  0.998  &  0.248  &  --  &  -- &  2.343  &  2.758  &  0.998  &  0.175  &  --  &  -- \\ \cline{3-21}
 &  & $\omega=1.0$ &  2.130  &  6.279  &  0.973  &  0.278  &  --  &  -- &  2.685  &  3.937  &  0.974  &  0.175  &  --  &  -- &  2.319  &  2.764  &  0.974  &  0.124  &  --  &  -- \\ \cline{3-21}
 &  & $\omega=1.5$ &  2.102  &  6.309  &  0.931  &  0.227  &  --  &  -- &  2.672  &  3.944  &  0.929  &  0.143  &  --  &  -- &  2.314  &  2.768  &  0.933  &  0.101  &  --  &  -- \\ \hline
\end{tabular}
}
\label{tab5}
\end{center}
{\footnotesize
Left side:\ sample size is $n=200$; Center:\ sample size is $n=500$; Right side:\ sample size is $n=1000$;\\
Logit:\ ordinary logistic regression; CBPS:\ Covariate balancing propensity score\cite{Im2014}; RCAL:\ Calibrating method proposed by Tan\cite{Ta2020}; BLPS:\ Bayesian method proposed by Saarela et al.\cite{Sa2015}; BRCAL:\ Proposed method;\\
{\bf (c)}:\ ``Moderate complexity and good overlap,'' {\bf (d)}:\ ``Moderate complexity and poor overlap.''
}
\end{table}

\begin{table}[htbp]
\begin{center}
\caption{Summary of estimated causal effects under the scenario {\bf (e)} and {\bf (f)} (``Predictor" situation):\ Bias, root mean squared error (RMSE), 95\% coverage probability (95\%CP), average length of CI (AvL), and bias and RMSE ratio (BR/RR) compared to the bias/RMSE of our proposed method of the IPW estimator in 10,000 iterations were summarized for each scenario (in ``Scenario" column) and method (in the ``Method'' column). RCAL:\ CV was used for the CV described in Section 4.2. BRCAL:\ PCIC was used for the PCIC to select the learning late $\omega$ from candidates: $(0.2, 0.5, 1.0, 1.5)$ (see Section 4.2).}
\scalebox{0.63}{
\begin{tabular}{ccc|cccccc|cccccc|cccccc}\hline
{\bf Scenario}&\multicolumn{2}{c|}{\bf Method}&\begin{tabular}{c}{\bf Bias}\\$(\times 10^{-2})$\end{tabular}&\begin{tabular}{c}{\bf RMSE}\\$(\times 10^{-2})$\end{tabular}&{\bf 95\%CP}&{\bf AvL}&{\bf BR}&{\bf RR}&\begin{tabular}{c}{\bf Bias}\\$(\times 10^{-2})$\end{tabular}&\begin{tabular}{c}{\bf RMSE}\\$(\times 10^{-2})$\end{tabular}&{\bf 95\%CP}&{\bf AvL}&{\bf BR}&{\bf RR}&\begin{tabular}{c}{\bf Bias}\\$(\times 10^{-2})$\end{tabular}&\begin{tabular}{c}{\bf RMSE}\\$(\times 10^{-2})$\end{tabular}&{\bf 95\%CP}&{\bf AvL}&{\bf BR}&{\bf RR}\\
\hline
(e) & \multicolumn{2}{c|}{Logit} &  3.150  &  8.186  &  0.910  &  0.266  &  1.234  &  1.044  &  4.181  &  5.107  &  0.935  &  0.181  &  1.185  &  1.030  &  4.084  &  3.555  &  0.942  &  0.132  &  1.209  &  1.018 \\ \cline{2-21} 
 & \multicolumn{2}{c|}{CBPS} &  2.509  &  7.922  &  0.914  &  0.278  &  0.983  &  1.010  &  3.515  &  4.981  &  0.925  &  0.177  &  0.996  &  1.005  &  3.369  &  3.502  &  0.936  &  0.129  &  0.998  &  1.002 \\ \cline{2-21}
 & RCAL & CV &  3.367  &  6.785  &  0.886  &  0.209  &  1.319  &  0.865  &  5.009  &  5.329  &  0.827  &  0.132  &  1.419  &  1.075  &  4.810  &  3.915  &  0.788  &  0.093  &  1.424  &  1.121 \\ \cline{3-21}
 &  & $\lambda=0.01$ &  10.530  &  14.410  &  0.728  &  3.457  &  --  &  --  &  7.313  &  6.792  &  0.743  &  0.133  &  --  &  --  &  5.235  &  4.102  &  0.766  &  0.093  &  --  &  --  \\ \cline{3-21}
 &  & $\lambda=0.05$ &  2.902  &  6.812  &  0.886  &  0.210  &  --  &  --  &  2.740  &  3.937  &  0.912  &  0.132  &  --  &  --  &  2.412  &  2.722  &  0.916  &  0.094  &  --  &  -- 
\\ \cline{3-21}
 &  & $\lambda=0.10$ &  3.353  &  5.514  &  0.938  &  0.207  &  --  &  --  &  3.984  &  3.452  &  0.940  &  0.130  &  --  &  --  &  3.997  &  2.454  &  0.940  &  0.092  &  --  &  -- \\ \cline{2-21}
 & \multicolumn{2}{c|}{BLPS} &  3.167  &  8.985  &  0.924  &  0.304  &  --  &  --  &  4.235  &  5.321  &  0.925  &  0.182  &  --  &  --  &  4.116  &  3.629  &  0.924  &  0.127  &  --  &  -- \\ \cline{2-21}
 & BRCAL & PCIC &  2.553  &  7.840  &  0.862  &  0.235  &  Ref.  &  Ref.  &  3.529  &  4.958  &  0.861  &  0.147  &  Ref.  &  Ref.  &  3.377  &  3.494  &  0.864  &  0.104  &  Ref.  &  Ref. \\ \cline{2-21}
 &  & $\omega=0.2$ &  2.798  &  7.397  &  1.000  &  0.620  &  --  &  --  &  3.565  &  4.824  &  1.000  &  0.392  &  --  &  --  &  3.402  &  3.438  &  1.000  &  0.277  &  --  &  -- \\ \cline{3-21}
 &  & $\omega=0.5$ &  2.639  &  7.710  &  0.988  &  0.391  &  --  &  --  &  3.541  &  4.917  &  0.988  &  0.248  &  --  &  --  &  3.387  &  3.478  &  0.987  &  0.175  &  --  &  -- \\ \cline{3-21}
 &  & $\omega=1.0$ &  2.576  &  7.817  &  0.921  &  0.276  &  --  &  --  &  3.526  &  4.949  &  0.926  &  0.175  &  --  &  --  &  3.381  &  3.489  &  0.925  &  0.124  &  --  &  -- \\ \cline{3-21}
 &  & $\omega=1.5$ &  2.548  &  7.851  &  0.852  &  0.226  &  --  &  --  &  3.527  &  4.959  &  0.854  &  0.143  &  --  &  --  &  3.378  &  3.494  &  0.856  &  0.101  &  --  &  -- \\ 
   \hline

(f) & \multicolumn{2}{c|}{Logit} &  7.391  &  8.060  &  0.920  &  0.268  &  1.279  &  1.021  &  8.451  &  5.041  &  0.932  &  0.179  &  1.286  &  1.013  &  8.031  &  3.595  &  0.931  &  0.129  &  1.323  &  1.013 \\ \cline{2-21}
 & \multicolumn{2}{c|}{CBPS} &  5.781  &  7.969  &  0.929  &  0.291  &  1.000  &  1.010  &  6.563  &  4.998  &  0.933  &  0.184  &  0.998  &  1.005  &  6.066  &  3.558  &  0.938  &  0.133  &  1.000  &  1.002 \\ \cline{2-21}
 & RCAL & CV &  6.878  &  7.447  &  0.857  &  0.208  &  1.190  &  0.944  &  11.817  &  6.799  &  NA  &  NA  &  1.797  &  1.366  &  12.872  &  5.486  &  --  &  --  &  2.121  &  1.545 
\\ \cline{3-21}
 &  & $\lambda=0.01$ &  19.805  &  19.236  &  0.721  &  292.898  &  --  &  --  &  17.018  &  9.565  &  0.640  &  0.161  &  --  &  --  &  14.547  &  6.070  &  0.626  &  0.100  &  --  &  -- \\ \cline{3-21}
 &  & $\lambda=0.05$ &  8.547  &  8.995  &  0.811  &  0.271  &  --  &  --  &  7.219  &  4.784  &  0.836  &  0.132  &  --  &  --  &  6.261  &  3.312  &  0.841  &  0.093  &  --  &  --\\ \cline{3-21}
 &  & $\lambda=0.10$ &  4.965  &  5.891  &  0.918  &  0.208  &  --  &  --  &  5.866  &  3.702  &  0.922  &  0.131  &  --  &  --  &  5.656  &  2.681  &  0.912  &  0.093  &  --  &  -- \\ \cline{2-21}
 & \multicolumn{2}{c|}{BLPS} &  7.685  &  8.790  &  0.931  &  0.304  &  --  &  --  &  8.588  &  5.214  &  0.924  &  0.182  &  --  &  --  &  8.085  &  3.651  &  0.921  &  0.127  &  --  &  -- \\ \cline{2-21}
 & BRCAL & PCIC &  5.780  &  7.892  &  0.856  &  0.230  &  Ref.  &  Ref.  &  6.574  &  4.976  &  0.858  &  0.146  &  Ref.  &  Ref.  &  6.068  &  3.550  &  0.856  &  0.104  &  Ref.  &  Ref. \\ \cline{3-21}
 &  & $\omega=0.2$ &  5.758  &  7.453  &  1.000  &  0.619  &  --  &  --  &  6.604  &  4.849  &  1.000  &  0.391  &  --  &  --  &  6.089  &  3.505  &  1.000  &  0.276  &  --  &  -- \\ \cline{3-21}
 &  & $\omega=0.5$ &  5.777  &  7.756  &  0.986  &  0.391  &  --  &  --  &  6.588  &  4.937  &  0.986  &  0.247  &  --  &  --  &  6.075  &  3.537  &  0.987  &  0.175  &  --  &  -- \\ \cline{3-21}
 &  & $\omega=1.0$ &  5.786  &  7.862  &  0.920  &  0.276  &  --  &  --  &  6.583  &  4.968  &  0.921  &  0.175  &  --  &  --  &  6.073  &  3.547  &  0.918  &  0.123  &  --  &  -- \\ \cline{3-21}
 &  & $\omega=1.5$ &  5.783  &  7.894  &  0.850  &  0.225  &  --  &  --  &  6.576  &  4.976  &  0.850  &  0.143  &  --  &  --  &  6.071  &  3.550  &  0.844  &  0.101  &  --  &  -- \\ \hline
\end{tabular}
}
\label{tab6}
\end{center}
{\footnotesize
Left side:\ sample size is $n=200$; Center:\ sample size is $n=500$; Right side:\ sample size is $n=1000$;\\
Logit:\ ordinary logistic regression; CBPS:\ Covariate balancing propensity score\cite{Im2014}; RCAL:\ Calibrating method proposed by Tan\cite{Ta2020}; BLPS:\ Bayesian method proposed by Saarela et al.\cite{Sa2015}; BRCAL:\ Proposed method;\\
{\bf (e)}:\ ``Large complexity and good overlap,'' {\bf (f)}:\ ``Large complexity and poor overlap.''
}
\end{table}
\end{landscape}

\begin{figure}[H]
\begin{center}
\begin{tabular}{c}
\includegraphics[width=18cm]{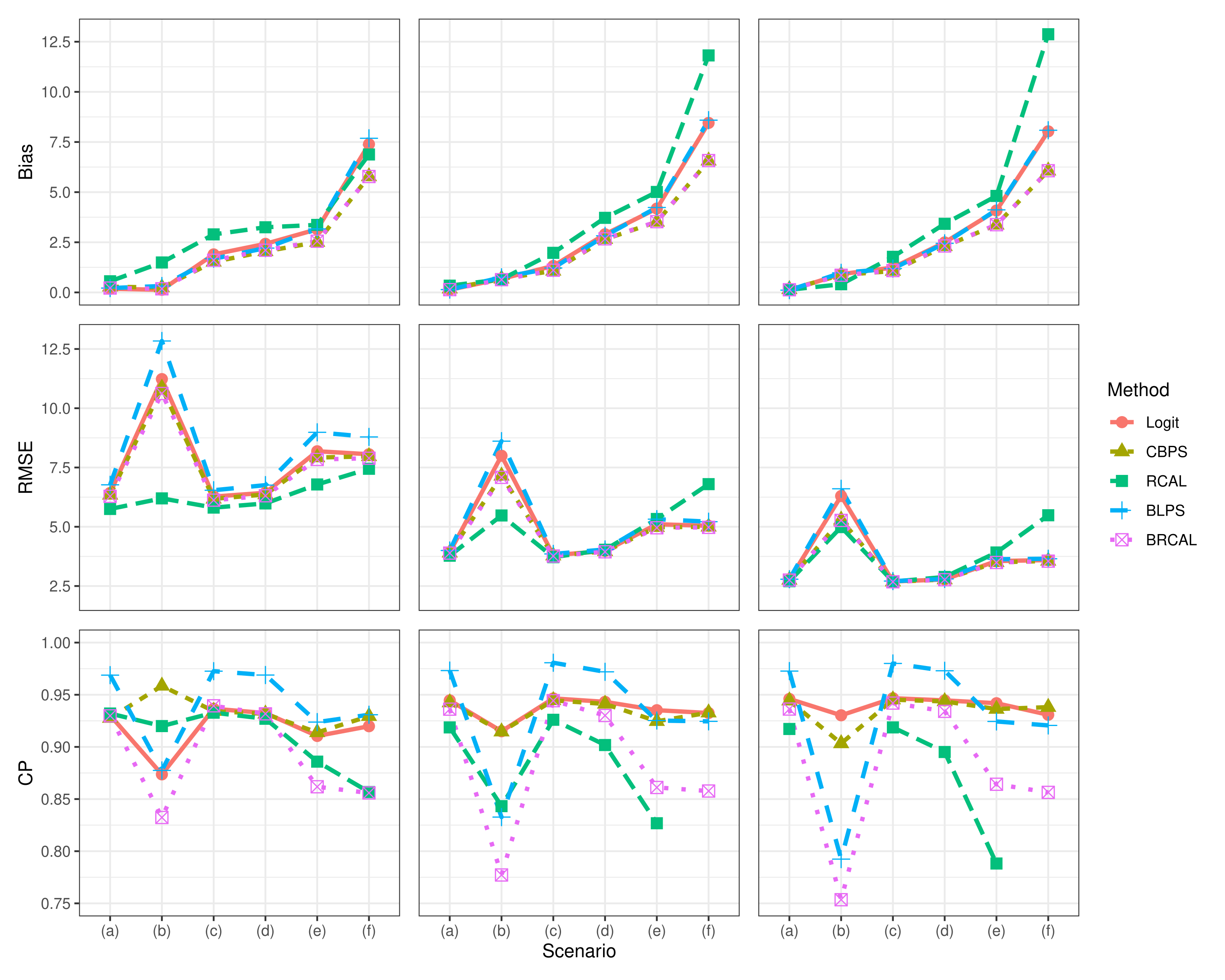}
\end{tabular}
\caption{Summary of estimated causal effects:\ Bias, root mean squared error (RMSE), and 95\% coverage probability (95\%CP) in 10,000 iterations were summarized under ``Predictor" situation. The left panel: $n=200$; the center panel: $n=500$; The right panel: $n=1000$. RCAL was used for the CV described in Section 4.2. BRCAL was used for the PCIC to select the learning rate $\omega$ from candidates: $(0.2, 0.5, 1.0, 1.5)$ (see Section 4.2).}
\label{fig2}
\end{center}
{\footnotesize
{\bf (a)}:\ ``Small complexity and good overlap,'' {\bf (b)}:\ ``Small complexity and poor overlap,'' {\bf (c)}:\ ``Moderate complexity and good overlap,'' {\bf (d)}:\ ``Moderate complexity and poor overlap,'' {\bf (e)}:\ ``Large complexity and good overlap,'' {\bf (f)}:\ ``Large complexity and poor overlap.''
}
\end{figure}

\newpage

\subsection{Covariate balance diagnosis}
The results of the covariate balancing diagnostics using the standardized mean differences (SMDs) are presented in Table \ref{tab7}. First, it is evident that confounding bias exists in this simulation dataset. In logistic regression, simple true propensity score settings, such as {\bf (a)} and {\bf (b)}, achieve good covariate balance. However, as the model complexity increases ({\bf (e)} and {\bf (f)}), this balance is no longer maintained. As expected, the CBPS ensures exact covariate balance across all settings. In contrast, the RCAL does not necessarily achieve as strong covariate balance as other methods. In some cases, the absolute value of the SMD exceeds 0.1, which may be undesirable in causal inference contexts.

Our proposed method does not achieve exact balance like the CBPS, but it maintains relatively good covariate balance. Combined with the simulation results in the main manuscript, these findings suggest that our proposed method performs similarly to the CBPS and outperforms the RCAL in terms of covariate balance. From a causal inference perspective, our proposed method follows the concept of a regularized approach by allowing some uncertainty in covariate balancing.

\begin{landscape}
\begin{table}[htbp]
\begin{center}
\caption{Summary of the standardized mean difference (SMD) under the scenario {\bf (a)}--{\bf (f)} (``Confounder" situation):\ SMD of the IPW estimator in 10,000 iterations were summarized for each scenario (in ``Scenario" column). RCAL:\ CV was used for the CV described in Section 4.2. BRCAL was used for the PCIC to select the learning rate $\omega$ from candidates: $(0.2, 0.5, 1.0, 1.5)$ (see Section 4.2).}
\scalebox{0.72}{
\begin{tabular}{cc|cccccc|cccccc|cccccc} \hline
{\bf Scenario} & {\bf Variables} & \multicolumn{18}{|c}{\bf SMD} \\
&   &  Before  &  Logit  &  CBPS  &  RCAL  &  BLPS  &  BRCAL  &  Before  &  Logit  &  CBPS  &  RCAL  &  BLPS  &  BRCAL  &  Before  &  Logit  &  CBPS  &  RCAL  &  BLPS  &  BRCAL 
\\ \hline
(a)  &  $X_1$  &  0.271  &  0.001  &  0.000  &  0.080  &  -0.012  &  0.014  &  0.268  &  0.000  &  0.000  &  0.018  &  -0.005  &  0.005  &  0.268  &  0.000  &  0.000  &  0.007  &  -0.002  &  0.002 \\ \cline{2-20}
  &  $X_2$  &  -0.318  &  0.000  &  0.000  &  -0.087  &  0.015  &  -0.015  &  -0.316  &  0.000  &  0.000  &  -0.018  &  0.006  &  -0.005  &  -0.315  &  0.000  &  0.000  &  -0.007  &  0.003  &  -0.002 \\ \cline{2-20}
  &  $X_3$  &  0.251  &  0.001  &  0.000  &  0.076  &  -0.011  &  0.013  &  0.251  &  0.000  &  0.000  &  0.018  &  -0.004  &  0.005  &  0.252  &  0.000  &  0.000  &  0.006  &  -0.002  &  0.002 \\ \cline{2-20}
  &  $X_4$  &  -0.169  &  0.000  &  0.000  &  -0.057  &  0.008  &  -0.010  &  -0.171  &  0.000  &  0.000  &  -0.015  &  0.003  &  -0.004  &  -0.169  &  0.000  &  0.000  &  -0.006  &  0.001  &  -0.002 \\ \hline
(b) &  $X_1$  &  0.433  &  0.005  &  0.000  &  0.085  &  -0.019  &  0.008  &  0.433  &  0.004  &  0.000  &  0.029  &  -0.006  &  0.003  &  0.432  &  0.004  &  0.000  &  0.015  &  -0.001  &  0.002 \\ \cline{2-20}
  &  $X_2$  &  -0.520  &  -0.003  &  0.000  &  -0.087  &  0.026  &  -0.008  &  -0.518  &  -0.002  &  0.000  &  -0.029  &  0.010  &  -0.003  &  -0.518  &  -0.002  &  0.000  &  -0.015  &  0.004  &  -0.002 \\ \cline{2-20}
  &  $X_3$  &  0.407  &  0.005  &  0.000  &  0.083  &  -0.018  &  0.008  &  0.406  &  0.003  &  0.000  &  0.028  &  -0.006  &  0.003  &  0.405  &  0.003  &  0.000  &  0.014  &  -0.001  &  0.002 \\ \cline{2-20}
  &  $X_4$  &  -0.269  &  -0.004  &  0.000  &  -0.073  &  0.011  &  -0.007  &  -0.269  &  -0.003  &  0.000  &  -0.026  &  0.003  &  -0.003  &  -0.270  &  -0.003  &  0.000  &  -0.013  &  0.000  &  -0.002 \\ \hline
(c)  &  $X_1$  &  0.289  &  0.015  &  0.000  &  0.121  &  0.002  &  0.014  &  0.286  &  0.015  &  0.000  &  0.072  &  0.010  &  0.005  &  0.285  &  0.015  &  0.000  &  0.061  &  0.012  &  0.002  \\ \cline{2-20}
  &  $X_2$  &  -0.027  &  -0.002  &  0.000  &  -0.013  &  0.000  &  -0.002  &  -0.030  &  -0.001  &  0.000  &  -0.008  &  -0.001  &  -0.001  &  -0.031  &  -0.001  &  0.000  &  -0.007  &  -0.001  &  -0.001  \\ \cline{2-20}
  &  $X_3$  &  0.345  &  0.012  &  0.000  &  0.121  &  -0.003  &  0.014  &  0.344  &  0.012  &  0.000  &  0.062  &  0.006  &  0.005  &  0.344  &  0.012  &  0.000  &  0.050  &  0.009  &  0.002  \\ \cline{2-20}
  &  $X_4$  &  -0.272  &  -0.002  &  0.000  &  -0.089  &  0.011  &  -0.013  &  -0.271  &  -0.002  &  0.000  &  -0.035  &  0.003  &  -0.005  &  -0.270  &  -0.002  &  0.000  &  -0.022  &  0.001  &  -0.002 \\ \hline
 (d) &  $X_1$  &  0.292  &  0.019  &  0.000  &  0.138  &  0.005  &  0.011  &  0.293  &  0.019  &  0.000  &  0.099  &  0.014  &  0.004  &  0.292  &  0.019  &  0.000  &  0.089  &  0.016  &  0.002   \\ \cline{2-20}
  &  $X_2$  &  -0.114  &  -0.003  &  0.000  &  -0.046  &  0.003  &  -0.006  &  -0.113  &  -0.003  &  0.000  &  -0.023  &  0.000  &  -0.003  &  -0.116  &  -0.003  &  0.000  &  -0.017  &  -0.001  &  -0.002   \\ \cline{2-20}
  &  $X_3$  &  0.392  &  0.013  &  0.000  &  0.128  &  -0.005  &  0.012  &  0.391  &  0.013  &  0.000  &  0.072  &  0.006  &  0.004  &  0.393  &  0.013  &  0.000  &  0.059  &  0.010  &  0.002   \\ \cline{2-20}
  &  $X_4$  &  -0.316  &  -0.002  &  0.000  &  -0.088  &  0.013  &  -0.011  &  -0.314  &  -0.002  &  0.000  &  -0.031  &  0.004  &  -0.004  &  -0.312  &  -0.002  &  0.000  &  -0.017  &  0.001  &  -0.002 \\ \hline
(e)  &  $X_1$  &  0.237  &  0.003  &  0.000  &  0.112  &  -0.010  &  0.007  &  0.235  &  0.002  &  0.000  &  0.065  &  -0.003  &  0.003  &  0.235  &  0.002  &  0.000  &  0.048  &  -0.001  &  0.002 \\ \cline{2-20}
  &  $X_2$  &  -0.643  &  -0.015  &  0.000  &  -0.061  &  0.020  &  -0.008  &  -0.639  &  -0.014  &  0.000  &  0.029  &  0.000  &  -0.003  &  -0.638  &  -0.013  &  0.000  &  0.054  &  -0.007  &  -0.002 \\ \cline{2-20}
  &  $X_3$  &  0.273  &  0.002  &  0.000  &  0.105  &  -0.013  &  0.007  &  0.275  &  0.001  &  0.000  &  0.048  &  -0.005  &  0.003  &  0.274  &  0.001  &  0.000  &  0.029  &  -0.003  &  0.002 \\ \cline{2-20}
  &  $X_4$  &  -0.453  &  -0.019  &  0.000  &  -0.044  &  0.005  &  -0.008  &  -0.449  &  -0.019  &  0.000  &  0.041  &  -0.009  &  -0.003  &  -0.448  &  -0.018  &  0.000  &  0.065  &  -0.014  &  -0.002 \\ \hline
(f)  &  $X_1$  &  0.146  &  0.007  &  0.000  &  0.116  &  -0.002  &  0.004  &  0.143  &  0.006  &  0.000  &  0.098  &  0.002  &  0.002  &  0.142  &  0.005  &  0.000  &  0.085  &  0.004  &  0.001 \\ \cline{2-20}
  &  $X_2$  &  -0.736  &  -0.052  &  0.000  &  -0.055  &  -0.011  &  -0.008  &  -0.737  &  -0.050  &  0.000  &  0.089  &  -0.034  &  -0.003  &  -0.736  &  -0.049  &  0.000  &  0.144  &  -0.041  &  -0.002 \\ \cline{2-20}
  &  $X_3$  &  0.225  &  0.001  &  0.000  &  0.129  &  -0.012  &  0.006  &  0.225  &  0.000  &  0.000  &  0.074  &  -0.006  &  0.003  &  0.224  &  -0.001  &  0.000  &  0.043  &  -0.004  &  0.001 \\ \cline{2-20}
  &  $X_4$  &  -0.576  &  -0.061  &  0.000  &  -0.039  &  -0.030  &  -0.007  &  -0.572  &  -0.058  &  0.000  &  0.100  &  -0.046  &  -0.003  &  -0.572  &  -0.058  &  0.000  &  0.153  &  -0.052  &  -0.002 \\ \hline
\end{tabular}
}
\label{tab7}
\end{center}
{\footnotesize
Left side:\ sample size is $n=200$; Center:\ sample size is $n=500$; Right side:\ sample size is $n=1000$;\\
Logit:\ ordinary logistic regression; CBPS:\ Covariate balancing propensity score\cite{Im2014}; RCAL:\ Calibrating method proposed by Tan\cite{Ta2020}; BLPS:\ Bayesian method proposed by Saarela et al.\cite{Sa2015}; BRCAL:\ Proposed method;\\
{\bf (a)}:\ ``Small complexity and good overlap,'' {\bf (b)}:\ ``Small complexity and poor overlap,'' {\bf (c)}:\ ``Moderate complexity and good overlap,'' {\bf (d)}:\ ``Moderate complexity and poor overlap,'' {\bf (e)}:\ ``Large complexity and good overlap,'' {\bf (f)}:\ ``Large complexity and poor overlap.''
}
\end{table}
\end{landscape}

\subsection{Sensitivity Analysis for Simulation Experiments}
Here, we conduct a sensitivity analysis (SA) for the simulation experiment results presented in the main manuscript. Specifically, we vary the hyperparameter for the prior distribution of $\lambda$. We examine three scenarios as follows. Note that all scenarios share the same prior mean of $0.1$, which is the common upper bound for the SMD.
\begin{description}
\item[Shape:\ 0.1, Rate:\ 1] Variance becomes:\ $0.1$; more strict setting compared to reference setting.
\item[Shape:\ 0.01, Rate:\ 0.1] This setting is the same as the main simulation results in the main manuscript. Therefore, the expected value of simulation results in SA becomes similar as them. Variance becomes:\ $1$ (reference setting).
\item[Shape:\ 0.002, Rate:\ 0.02] Variance becomes:\ $5$; more optimistic setting compared to reference setting.
\end{description}

The SA results are presented in Table \ref{tab8}. In small sample sizes and simple propensity score model settings, a strict prior setting resulted in smaller bias; however, it also increased the variance of the posterior distribution, leading to wider confidence intervals (CIs). On the other hand, in other settings, such strict balancing does not appear to be necessary. Additionally, an optimistic prior setting does not introduce bias in the ATE estimates. In summary, we conclude that the effect of the prior setting for $\lambda$ on the ATE estimate is limited.

\newpage
\begin{landscape}
\begin{table}[htbp]
\begin{center}
\caption{Summary of estimated causal effects under the scenario {\bf (a)}--{\bf (f)} (``Confounder" situation):\ Bias, root mean squared error (RMSE), 95\% coverage probability (95\%CP), and average length of CI (AvL) of the IPW estimator in 10,000 iterations were summarized for each scenario (in ``Scenario" column). PCIC was used to select the learning rate $\omega$ from candidates: $(0.2, 0.5, 1.0, 1.5)$ (see Section 4.2).}
\scalebox{0.9}{
\begin{tabular}{ccc|cccc|cccc|cccc} \hline
{\bf Scenario} & {\bf Shape} & {\bf Rate} & {\bf Bias} & {\bf RMSE} & {\bf 95\%CP} & {\bf AvL}  & {\bf Bias} & {\bf RMSE} & {\bf 95\%CP} & {\bf AvL} & {\bf Bias} & {\bf RMSE} & {\bf 95\%CP} & {\bf AvL} \\ \hline
(a)  &  0.1  &  1  &  0.009  &  5.560  &  0.990  &  0.519  &  0.085  &  3.493  &  0.992  &  0.324  &  0.020  &  2.475  &  0.993  &  0.227  \\ \cline{2-15}
  &  0.01  &  0.1  &  0.041  &  5.597  &  0.959  &  0.259  &  0.047  &  3.505  &  0.960  &  0.146  &  0.020  &  2.478  &  0.965  &  0.104  \\ \cline{2-15}
  &  0.002  &  0.02  &  0.017  &  5.598  &  0.957  &  0.234  &  0.050  &  3.504  &  0.958  &  0.144  &  0.017  &  2.478  &  0.962  &  0.103 \\ \hline
(b)  &  0.1  &  1  &  0.285  &  6.237  &  0.941  &  0.262  &  0.157  &  3.871  &  0.945  &  0.155  &  0.067  &  2.730  &  0.952  &  0.112 \\ \cline{2-15}
  &  0.01  &  0.1  &  0.299  &  6.235  &  0.931  &  0.228  &  0.165  &  3.871  &  0.936  &  0.146  &  0.067  &  2.730  &  0.943  &  0.105 \\ \cline{2-15}
  &  0.002  &  0.02  &  0.296  &  6.235  &  0.931  &  0.228  &  0.167  &  3.870  &  0.934  &  0.146  &  0.069  &  2.729  &  0.943  &  0.105 \\ \hline
(c)  &  0.1  &  1  &  0.041  &  5.551  &  0.989  &  0.492  &  0.362  &  3.428  &  0.993  &  0.313  &  0.147  &  2.413  &  0.995  &  0.222 \\ \cline{2-15}
  &  0.01  &  0.1  &  0.236  &  5.582  &  0.960  &  0.253  &  0.499  &  3.435  &  0.964  &  0.146  &  0.204  &  2.416  &  0.968  &  0.105\\ \cline{2-15} 
  &  0.002  &  0.02  &  0.263  &  5.578  &  0.958  &  0.233  &  0.487  &  3.436  &  0.963  &  0.144  &  0.200  &  2.415  &  0.967  &  0.103 \\ \hline
(d)  &  0.1  &  1  &  0.146  &  5.605  &  0.980  &  0.425  &  0.766  &  3.515  &  0.982  &  0.243  &  0.544  &  2.481  &  0.982  &  0.164 \\ \cline{2-15}
  &  0.01  &  0.1  &  0.376  &  5.631  &  0.958  &  0.236  &  0.843  &  3.519  &  0.960  &  0.145  &  0.564  &  2.483  &  0.962  &  0.104 \\ \cline{2-15}
  &  0.002  &  0.02  &  0.375  &  5.627  &  0.956  &  0.228  &  0.834  &  3.518  &  0.959  &  0.144  &  0.566  &  2.482  &  0.962  &  0.103 \\ \hline
(e)  & 0.1  &  1  &  1.231  &  6.410  &  0.941  &  0.266  &  1.309  &  4.013  &  0.939  &  0.157  &  1.168  &  2.812  &  0.948  &  0.114 \\ \cline{2-15}
  &  0.01  &  0.1  &  1.229  &  6.416  &  0.927  &  0.229  &  1.308  &  4.012  &  0.932  &  0.146  &  1.171  &  2.812  &  0.933  &  0.105 \\ \cline{2-15}
  &  0.002  &  0.02  &  1.226  &  6.414  &  0.925  &  0.228  &  1.309  &  4.012  &  0.929  &  0.146  &  1.167  &  2.811  &  0.932  &  0.105 \\ \hline
(f)  &  0.1  &  1  &  2.866  &  6.831  &  0.917  &  0.247  &  2.547  &  4.291  &  0.926  &  0.158  &  2.473  &  3.027  &  0.928  &  0.115 \\ \cline{2-15}
  &  0.01  &  0.1  &  2.885  &  6.830  &  0.908  &  0.231  &  2.547  &  4.289  &  0.916  &  0.148  &  2.468  &  3.026  &  0.915  &  0.107 \\ \cline{2-15}
  &  0.002  &  0.02  &  2.885  &  6.828  &  0.908  &  0.230  &  2.560  &  4.290  &  0.914  &  0.148  &  2.473  &  3.027  &  0.913  &  0.106 \\ \hline
\end{tabular}
}
\label{tab8}
\end{center}
{\footnotesize
Left side:\ sample size is $n=200$; Center:\ sample size is $n=500$; Right side:\ sample size is $n=1000$;\\
Logit:\ ordinary logistic regression; CBPS:\ Covariate balancing propensity score\cite{Im2014}; RCAL:\ Calibrating method proposed by Tan\cite{Ta2020}; BLPS:\ Bayesian method proposed by Saarela et al.\cite{Sa2015}; BRCAL:\ Proposed method;\\
{\bf (a)}:\ ``Small complexity and good overlap,'' {\bf (b)}:\ ``Small complexity and poor overlap,'' {\bf (c)}:\ ``Moderate complexity and good overlap,'' {\bf (d)}:\ ``Moderate complexity and poor overlap,'' {\bf (e)}:\ ``Large complexity and good overlap,'' {\bf (f)}:\ ``Large complexity and poor overlap.''
}
\end{table}
\end{landscape}

\newpage
\section{Additional Materials for Real Data Analysis}
\label{appC}
In the real data analysis, we use the dataset known as the ``Whitehall" dataset. Royston et al. \cite{Ro1999} investigated the relationship between mortality and various factors, including socioeconomic status, using 10-year follow-up data from the Whitehall I study. This prospective cohort study included 18,403 male British civil servants working in London, excluding those in the Diplomatic Service and British Council. After excluding four participants with implausible systolic blood pressure readings and 106 with less than 10 years of follow-up, 17,260 men remained, of whom 1,670 (9.7\%) died. Assuming non-informative censoring, logistic regression was used to estimate the odds ratio of death within 10 years in Royston et al. \cite{Ro1999}. In their analysis, cigarette smoking and systolic blood pressure were treated as continuous risk factors, while age, plasma cholesterol concentration, and Civil Service grade were included as confounders.

\begin{figure}[htbp]
\begin{center}
\begin{tabular}{c}
\includegraphics[width=15cm]{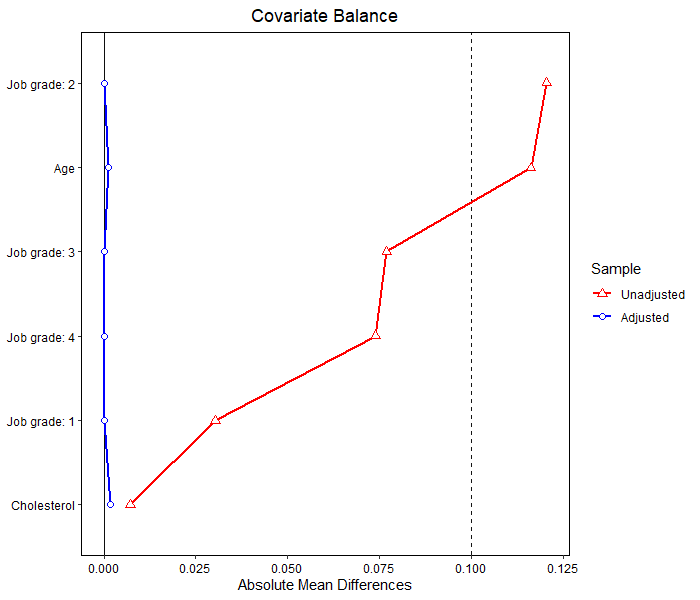}
\end{tabular}\caption{SMD of confounders in Whitehall dataset.}
\label{fig_demo}
\end{center}
\end{figure}

Based on their manuscript, the same subjects are considered, with smoking status as the exposure variable and all-cause death within 10 years as the outcome. The baseline characteristics and standardized mean differences (SMDs) before and after confounding adjustment using inverse probability weighting (IPW) with ordinary logistic regression are summarized in Table \ref{tab_demo}. As expected, the SMDs are smaller after adjustment than before. The absolute values of the SMDs are also plotted in Figure \ref{fig_demo}. Combining these results with those in Section B.3, we expect that covariate balance is also achieved in our proposed method.

\begin{table}[htbp]
\begin{center}
\caption{Demographic of Whitehall dataset}
\begin{tabular}{cc|ccc|cc}\hline
 & & {\bf Non-smoking} & {\bf Smoking} & {\bf Overall} & \multicolumn{2}{|c}{\bf SMD}\\
& & $N=10103$ & $N=7157$ & $N=17260$ & Before & After \\\hline
Age & & 51.3 (6.27) & 52.0 (6.35) & 51.6 (6.31) & -0.116 & -0.001 \\\hline
Cholesterol & & 5.11 (1.20) & 5.10 (1.19) & 5.11 (1.20) & 0.007 & -0.002 \\\hline
Job grade & 1 & 682 (6.8\%) & 266 (3.7\%) & 948 (5.5\%) & 0.137 & 0.000 \\
& 2 & 7539 (74.6\%) & 4478 (62.6\%) & 12017 (69.6\%) & 0.262 & 0.000 \\
& 3 & 1265 (12.5\%) & 1447 (20.2\%) & 2712 (15.7\%) & -0.209 & 0.000 \\
& 4 & 617 (6.1\%) & 966 (13.5\%) & 1583 (9.2\%) & -0.250 & 0.000 \\\hline
\end{tabular}
\label{tab_demo}
\end{center}
{\footnotesize
For continuous variables:\ mean (SD); for categorical variables:\ number of subjects (\%);\\
Before:\ Before IP weighting; After:\ After IP weighting.
}
\end{table}

\newpage
\section{Proposed Method under some Covariate Missing}
\label{appD}
In this section, we address the scenario where one covariate is missing, with the missing mechanism being Missing At Random (MAR)\cite{Li2019}. In this context, we explore the covariate balancing-based estimating process for the missing covariate, applying the same process as discussed in the main manuscript. Here, our focus is solely on outlining the construction of the estimation process. To simplify the discussion, we consider the situation where only one covariate is potentially missing.

Let $R_{i}$ denote the missing indicator: $R_{i}=1$ means observed and $R_{i}=0$ means missing. Also, let $\bld{X}_{i}=(R_{i}X_{1i},\bld{X}_{2i})\in\mathbb{R}^{1}\times\mathbb{R}^{p-1}$ denote the (observed) confounders. Under this setting, the observed data is described by the five pairs: $(A_{i},R_{i},Y_{i},R_{i}X_{i1},\bld{X}_{2i})$. The missing mechanism is denoted as $e^{R}_{i}(\bld{X}_{2i})={\rm Pr}(R=1|\bld{X}_{2i})$. Additionally, the propensity score is redefined as $e^{A}_{i}(R_{i},\bld{X}_{i})={\rm Pr}(A=1|R_{i},\bld{X}_{i})$. Under the following two independence assumptions, the IPW estimator for the treatment group (i.e., $\theta_{1}$), denoted by
$$
\hat{\theta}_{1}=\frac{\sum_{i=1}^{n}\frac{R_{i}A_{i}Y_{i}}{e^{R}_{i}e^{A}_{i}}}{\sum_{i=1}^{n}\frac{R_{i}A_{i}}{e^{R}_{i}e^{A}_{i}}},
$$
becomes a consistent estimator:
\begin{align}
(Y_{1},Y_{0})\indep (A,R)|\bld{X},\ \ \ R\indep X_{1}|\bld{X}_{2}\label{indep_mis}.
\end{align}
The former assumption is a straightforward extension of the strongly ignorable treatment assignment, and the latter is considered the MAR assumption (i.e., the potentially missing variable is independent of the missing indicator). Thus,
$$
\frac{\sum_{i=1}^{n}\frac{R_{i}A_{i}Y_{i}}{e^{R}_{i}e^{A}_{i}}}{\sum_{i=1}^{n}\frac{R_{i}A_{i}}{e^{R}_{i}e^{A}_{i}}}\stackrel{P}{\to}\frac{{\rm E}\left[\frac{RAY}{e^{R}e^{A}}\right]}{{\rm E}\left[\frac{RA}{e^{R}e^{A}}\right]}=\frac{{\rm E}\left[{\rm E}\left[\frac{RA}{e^{R}e^{A}}|\bld{X}\right]{\rm E}[Y_{1}|\bld{X}]\right]}{{\rm E}\left[{\rm E}\left[\frac{RA}{e^{R}e^{A}}|\bld{X}\right]\right]},
$$
and
$$
{\rm E}\left[\frac{RA}{e^{R}e^{A}}|\bld{X}\right]={\rm E}\left[\frac{RA}{{\rm Pr}(R=1,A=1|\bld{X})}|\bld{X}\right]=1.
$$
Therefore, from (\ref{indep_mis}), a similar moment conditions in the main manuscript can be considered. For instance, we can consider the (\ref{eq1}), replaced with following loss functions:
\begin{align*}
\ell_{R}(\bld{\gamma})&=\sum_{i=1}^{n}\left[R_{i}\exp\left\{-\bld{\gamma}^{\top}g_{1}(\bld{X}_{i})\right\}+(1-R_{i})\bld{\gamma}^{\top}g_{1}\right]\\
\ell_{A}(\bld{\alpha})&=\sum_{i=1}^{n}\left[A_{i}\exp\left\{-\bld{\alpha}^{\top}R_{i}g_{2}(\bld{X}_{i})\right\}+(1-A_{i})\bld{\alpha}^{\top}R_{i}g_{2}+(1-A_{i})\exp\left\{\bld{\alpha}^{\top}R_{i}g_{2}\right\}-A_{i}\bld{\alpha}^{\top}R_{i}g_{2}\right].
\end{align*}
Using the loss functions, we can construct the similar Bayesian estimation procedure described in the main manuscript.
\end{document}